\documentclass{aa}

\usepackage{graphicx}
\usepackage{natbib}
\usepackage{enumerate}

\bibpunct{(}{)}{;}{a}{}{,} % to follow A\&A style

\begin{document}

%macros
\newcommand{\simgt}{\lower.5ex\hbox{$\;\buildrel>\over\sim\;$}}
\newcommand{\simlt}{\lower.5ex\hbox{$\;\buildrel<\over\sim\;$}}
\newcommand{\hst}{{\sl HST}}
\newcommand{\spitzer}{{\sl Spitzer}}
\newcommand{\lya}{{Ly$\alpha$}}                      
\newcommand{\bra}{{Br$\alpha$}}                      
\newcommand{\brg}{{Br$\gamma$}}                      
\newcommand{\ha}{{H$\alpha$}}                      
\newcommand{\htwo}{{H$_2$}}                      
\newcommand{\hb}{{H$\beta$}}             
\newcommand{\zsun}{{$Z_\odot$}}                      
\newcommand{\msun}{{$M_\odot$}}                      
\newcommand{\lsun}{{$L_\odot$}}                      
\newcommand{\av}{{$A_V$}}
\newcommand{\vi}{{$V-I$}}
\newcommand{\hii}{{H{\sc ii}}}
\newcommand{\hi}{{H{\sc i}}}
\newcommand{\Ne}{{N$_e$}}
\newcommand{\magsq}{mag\,arcsec$^{-2}$}
\newcommand{\micron}{\,$\mu$m}
\newcommand{\nnn}{NGC\,2782}
\newcommand{\coone}{$^{12}$CO(1--0)}
\newcommand{\cotwo}{$^{12}$CO(2--1)}
\newcommand{\twelveco}{$^{12}$CO}
\newcommand{\vsys}{$V_{sys}$}
\newcommand{\vrot}{$V_{\rm rot}$}
\newcommand{\vcirc}{$V_{\rm circ}$}
\newcommand{\kms}{km\,s$^{-1}$}
\newcommand{\kmskpc}{km\,s$^{-1}$\,kpc$^{-1}$}

% spectral lines
\newcommand{\nii}{\ensuremath{\mathrm{[N\,II]}}}
\newcommand{\oiii}{\ensuremath{\mathrm{[O\,III]}}}
\newcommand{\oii}{\ensuremath{\mathrm{[O\,II]}}}
\newcommand{\sii}{\ensuremath{\mathrm{[S\,II]}}}

\newcommand{\Ho}{\ensuremath{\mathrm{H}_0}}
\newcommand{\Msun}{\ensuremath{~\mathrm{M}_\odot}}
\newcommand{\Lsun}{\ensuremath{~\mathrm{L}_\odot}}
\newcommand{\LBsun}{\ensuremath{~\mathrm{L}_{B\odot}}}
\newcommand{\AV}{\ensuremath{\mathrm{A}_V}}
\newcommand{\QH}{\ensuremath{Q(\mathrm{H})}}
% BHs
\newcommand{\bh}{\ensuremath{\mathrm{BH}}}
\newcommand{\mbh}{\ensuremath{M_\mathrm{BH}}}
\newcommand{\mdyn}{\ensuremath{M_\mathrm{dyn}}}
\newcommand{\lsph}{\ensuremath{L_\mathrm{bul}}}
\newcommand{\LV}{\ensuremath{L_\mathrm{V}}}
\newcommand{\lopt}{\ensuremath{L_\mathrm{opt}}}

% Units

%
%%
\newcommand{\mlr}{\ensuremath{\Upsilon}}
\newcommand{\I}{\ensuremath{i}}
\newcommand{\Th}{\ensuremath{\theta}}
\newcommand{\B}{\ensuremath{b}}
\newcommand{\So}{\ensuremath{\mathrm{s}_\circ}}
\newcommand{\Vsys}{\ensuremath{V_\mathrm{sys}}}
\newcommand{\chisq}{\ensuremath{\chi^2}}
\newcommand{\chisqr}{\ensuremath{\chi^2_\mathrm{red}}}
\newcommand{\chisqc}{\ensuremath{\chi^2_\mathrm{c}}}

\title{Molecular Gas in NUclei of GAlaxies (NUGA): \\ 
IX. The decoupled  bars and gas inflow in NGC\,2782
\thanks{Based on observations
carried out with the IRAM Plateau de Bure Interferometer. IRAM is supported by the
INSU/CNRS (France), MPG (Germany), and IGN (Spain).}
}

\author{L.K. Hunt \inst{1}
\and F. Combes \inst{2}
\and S. Garc\'ia-Burillo \inst{3}
\and E. Schinnerer \inst{4}
\and M. Krips \inst{5}
\and A.J. Baker \inst{6}
\and F. Boone \inst{2}
\and A. Eckart \inst{7}
\and S. L\'eon \inst{8}
\and R. Neri \inst{9}
\and L.J. Tacconi \inst{10}
   }

\offprints{{\tt hunt@arcetri.astro.it}}

\institute{
INAF-Istituto di Radioastronomia/Sez. Firenze, Largo Enrico Fermi 5, 50125 Firenze, Italy;
\email{hunt@arcetri.astro.it}
\and Observatoire de Paris, LERMA, 61 Av. de l'Observatoire, F-75014, Paris, France 
\and Observatorio Astron\'omico Nacional (OAN) - Observatorio de Madrid,
    C/ Alfonso XII, 3, 28014 Madrid, Spain
\and Max-Planck-Institut f\"ur Astronomie, K\"onigstuhl 17, D-69117 Heidelberg, Germany
\and Harvard-Smithsonian Center for Astrophysics, SMA, 645 N. A'ohoku Pl., Hilo, HI 96720 USA
\and Department of Physics and Astronomy, Rutgers, State University of New Jersey, 136 Frelinghuysen Road,
Piscataway, NJ 08854, USA
\and I. Physikalisches Institut, Universit\"at zu K\"oln, Z\"ulpicherstrasse 77, 
50937-K\"oln, Germany
\and IRAM-Pico Veleta Observatory, Avenida Divina Pastora 7, Local 20 E-18012 Granada, Spain
\and IRAM-Institut de Radio Astronomie Millim\'etrique, 300 Rue de la Piscine,
38406-St.Mt.d`H\`eres, France
\and Max-Planck-Institut f\"ur extraterrestrische Physik, Postfach 1312, D-85741 Garching, Germany
      }

\date{Received ; accepted}

\abstract{
We present CO(1-0) and CO(2-1) maps of the starburst/Seyfert\,1 galaxy \nnn\
obtained with the IRAM interferometer, at 2\farcs1$\times$1\farcs5
and 0\farcs7$\times$0\farcs6\ resolution
respectively. The CO emission is aligned along the stellar nuclear bar 
of radius $\sim$1\,kpc, configured in an elongated structure with two spiral arms
at high pitch angle $\sim90^\circ$. 
At the extremity of the nuclear bar, the CO changes direction to
trace two more extended spiral features at a lower pitch angle.
These are the beginning of two straight dust lanes, which are aligned
parallel to an oval distortion, reminiscent of a primary bar,
almost perpendicular to the nuclear one. 
The two embedded bars appear in Spitzer IRAC near-infrared images,
and HST color images, although highly obscured by dust in the latter.
We compute the torques exerted by the stellar bars on the gas,
and find systematically negative average torques down to the resolution limit of the images,
providing evidence of gas inflow tantalizingly close to the nucleus of \nnn.
We propose a dynamical scenario based on numerical simulations
to interpret coherently the radio, optical,
and molecular gas features in the center of the galaxy. 
Star formation is occurring in a partial ring 
at $\sim$\,1.3\,kpc radius corresponding to the Inner Lindblad Resonance (ILR) 
of the primary bar; 
this ring-like structure encircles the nuclear bar, and is studded with
\ha\ emission. 
The gas traced by CO emission is driven inward
by the gravity torques of the decoupled nuclear bar, since most of it is inside
its corotation. 
N-body simulations, including gas dissipation, predict the secondary bar decoupling,
the formation of the elongated ring at the $\sim$1\,kpc-radius ILR of the primary bar,
and the gas inflow to the ILR of the nuclear bar at a radius of $\sim$200-300\,pc. 
The presence of molecular gas inside the ILR of the primary bar,
transported by a second nuclear bar, is a potential ``smoking gun'';
the gas there is certainly fueling the central starburst, 
and in a second step could fuel directly the AGN.
}

\authorrunning{Hunt et al.} 
\titlerunning{NUGA: IX. NGC\,2782}

\maketitle

\section{Introduction}

Molecular gas is the dominant gas component in the inner regions of
spiral galaxies, making
CO lines unique tracers of nuclear gas dynamics.
As such, they are also a powerful diagnostic for identifying 
how active nuclei (AGN) are fueled.
To feed an AGN through accretion, there must be an adequate supply
of gas
whose angular momentum has been sufficiently reduced to enable inflow
within the small spatial scales surrounding the black hole.
Although there is rarely a lack of circumnuclear fuel, 
it is not yet clear how angular momentum is removed
to enable nuclear accretion. 

To better understand how AGN are fed and maintained,
we have been conducting for several years now a high-resolution high-sensitivity
CO survey (NUGA, \citealt{santi03}) of galaxies 
at the IRAM Plateau de Bure Interferometer (PdBI).
Altogether we have observed 12 galaxies in two CO transitions
with up to four configurations of the array, so as to achieve the most sensitive
(typically $\sim2-4$ mJy\,beam$^{-1}$ in 10\,km\,s$^{-1}$ channels) and 
the highest resolution (1-2\,\arcsec) survey currently available.
The results of the NUGA survey so far have been surprising: there is no single
unambiguous circumnuclear molecular gas feature connected with the nuclear activity.
One- and two-armed instabilities \citep{santi03},
well-ordered rings and nuclear spirals \citep{francoise04},
circumnuclear asymmetries \citep{melanie05},
and large-scale bars \citep{fred07} are among
the variety of molecular gas morphologies revealed by our survey. 
Moreover, an analysis of the torques exerted by the
stellar gravitational potential on the molecular gas in four galaxies suggests
that the gas tends to be driven away from the
the AGN ($\simgt$50\,pc), rather than toward it \citep{santi05}.
Nevertheless, these dynamics do not correspond to the violent molecular outflows 
and superwinds predicted in AGN feedback models \citep[e.g.,][]{narayanan06,hopkins06}, 
because the observed velocities are much too small.

Much of the explanation of this variety of morphologies appears to be related
to timescales \citep{santi05}.
It is well established that large-scale bars transport gas
inward very efficiently \citep[e.g.,][]{combes85,sakamoto99}, 
and there is very little doubt that bars can drive powerful starbursts 
\citep{knapen02,jogee05}.
However, no clear correlation between bars and nuclear activity 
has yet been found \citep[e.g.,][]{mulchaey97}. 
This may be because the timescales for bar-induced gas inflow 
and AGN duty cycles are very different.
Bars drive inflow over timescales ($\simgt$300\,Myr) that are 
similar to the typical gas-consumption timescales of a few 
times $\sim10^8$\,yr found in nuclear starbursts \citep[e.g.,][]{jogee05}.
But AGN accretion-rate duty cycles are much shorter than this
($\sim$1-10\,Myr, \citealt{heckman04,hopkins06,king07}), and
there are several indications that active accretion
occurs only intermittently over the lifetime of a galaxy
\citep{ferrarese01,marecki03,janiuk04,hopkins06,king07}. 
The resulting implication is that most AGN are probably between active 
accretion episodes, and catching galaxies with nuclear accretion 
``switched on'' may be difficult.

In this paper, the ninth of the NUGA series,
we present observations that suggest that we have
found one of these potentially rare
AGN with possible gas inflow in the current epoch.
\nnn\ is an early-type spiral galaxy [SABa(rs)] with peculiar morphology.
In addition to a pronounced stellar tail or sheet $\sim$20\,kpc to the east, it shows 
three optical ripples \citep[e.g.,][]{smith94} thought to be signatures of 
tidal interactions \citep{schweizer88}.
A massive H{\sc i} plume extends $\sim$54\,kpc to the northwest, 
and the neutral atomic gas in the inner disk is counterrotating
with respect to the gas motions in the outer regions \citep{smith91}.
The central regions of \nnn\ host a massive nuclear starburst, with a far-infrared
(FIR) luminosity of $2\times10^{10}$\msun, comparable to that in M\,82 \citep{devereux89}.
Three-dimensional optical spectroscopy \citep{yoshida99} shows evidence for a high-speed
ionized gas outflow, with the bipolar structure in the radio 
continuum indicative of a confined superbubble \citep{jogee98}.
In the outflow, there are also high-excitation extranuclear emission lines
thought to be due to shock heating \citep{boer92}. 

Until recently, it was thought that the outflow and energetics in \nnn\ were powered by
a starburst alone, but recent radio and X-ray observations reveal an optically-hidden AGN.
MERLIN and EVN/VLBI observations show a high-brightness-temperature
extremely compact ($\la$0\farcs05) radio source, unambiguous
evidence of an AGN \citep{melanie07}.
\nnn\ is also a Compton-thick X-ray source 
with a 6.4\,keV iron feature coming from its innermost regions \citep{zhang06}.
There is a bright unresolved X-ray core
and extended emission roughly coincident with the radio morphology in the
high-resolution ($\sim1$\arcsec) image by \citet{saikia94}.

The nuclear region of \nnn\ has been observed previously in the \coone\ line by 
\citet{ishuzuki94} with the Nobeyama Millimeter Array and by \citet{jogee99}
using the Owens Valley Radio Observatory.
We reobserved \nnn\ in \coone\ at PdBI with better spatial resolution 
and a sensitivity roughly five times that of previous observations, and for the
first time in the \cotwo\ line.
This enables a rigorous derivation of the torques acting on the molecular gas in \nnn,
and a quantitative assessment of the infall of material to the nucleus.
We first present our new observations in Section \ref{sec:obs},
together with our multiwavelength imaging dataset. 
The morphology and kinematics of the molecular gas are discussed in
Section \ref{sec:coresults},
and we describe the stellar structure in Section \ref{sec:stars}
and the starburst episode in Section \ref{sec:dust}.
We then derive the gravitational potential from the infrared
image, and infer the torques acting on the molecular gas
in Section \ref{sec:torques}.
Finally, we present numerical simulations which motivate our proposed
scenario of decoupled double bars in this galaxy.
The molecular gas in \nnn\ is apparently being driven inward by the nuclear bar, 
decoupled from the primary bar, since we detect azimuthally averaged torques 
which are negative down to the resolution limit of our images.

\section{Observations \label{sec:obs}}

We observed \nnn\ with the IRAM PdBI in the ABCD configuration of the array
between December 2001 and February 2003 in the \coone\ (115\,GHz) and the
\cotwo\ (230\,GHz) rotational transitions. The PdBI receiver characteristics,
the observing procedures, and the image reconstruction are the same as those 
described in \citet{santi03}.  
The quasar 3C273 was used for bandpass calibration and the
quasar 0923$+$392 was used to calibrate both the phase and the atmospheric variations. 
We used uniform weighting to generate 2--1 maps with a
field of view of 21\arcsec\ and natural weighting to produce the 1--0 maps
with a field of view of 42\arcsec.  
Such a procedure maximizes the flux recovered in CO(1--0) and optimizes the
spatial resolution in CO(2--1).

The {\it rms} noise $\sigma$ in
10\,km\,s$^{-1}$ wide velocity channels is 2.0\,mJy\,beam$^{-1}$ and
5.2\,mJy\,beam$^{-1}$, with beam sizes of 2\farcs1$\times$1\farcs5 and
0\farcs7$\times$0\farcs6 at 115 and 230\,GHz, respectively.  At a level
of $\sim3\sigma$, no 3\,mm (1\,mm) continuum is detected toward \nnn\ to a
level of 1\,mJy\,beam$^{-1}$ (3\,mJy\,beam$^{-1}$).  The conversion factors
between Jy\,beam$^{-1}$ and K are 30\,K\,Jy$^{-1}$\,beam at 115\,GHz, and
58\,K\,Jy$^{-1}$\,beam at 230\,GHz.  By default, all velocities are referred to
the heliocentric recession velocity $v_0$\,=\,2545\,km\,s$^{-1}$ and $(\Delta
\alpha, \Delta \delta)$ offsets are relative to the phase tracking center of
the observations (RA$_{2000}$, Dec$_{2000}$)=(09$^h$14$^m$05.08$^s$,
40$^d$06$^m$49.4$^s$). 
The displayed maps are not corrected for primary beam attenuation.  

We will assume a distance
to \nnn\ of $D=$35\,Mpc, which is derived from the local velocity field model
given in \citet{mould00} and a Hubble constant
$H_0=73$\,km\,s$^{-1}$\,Mpc$^{-1}$.  
At this distance 1\arcsec\ corresponds to 171\,pc.

\subsection{Optical and infrared images \label{sec:otherdata}}

We retrieved \hst\ archival images of \nnn\ with WFPC2 in
the F555W and F814W filters.
Cosmic rays were eliminated, and the images were calibrated 
and converted to $V$ and $I$ as described in \citet{holtzman95}.
We performed an astrometric calibration using stars
from the U.S. Naval Observatory Astrometric Catalog
B1.0 (USNO-B1.0).
Five stars from this catalogue appear in the 1600$\times$1600 WFPC2
image of \nnn\ and they were used to derive
the astrometric solution with {\it imwcs} in the WCSTools 
package\footnote{Available from\\ {\it http://tdc-www.harvard.edu/software/wcstools}}.
The solution has an rms uncertainty of 0\farcs24, or 2.4 WFPC2 (mosaic) pixels,
and differs from the original \hst\ one by $\simgt$1\arcsec.
We made \vi\ color images by
converting the flux units to magnitudes and subtracting the two magnitude images.
%An extinction image was derived from the \vi\ color image
%by assuming an unreddened color of \vi\,=\,1.03
%for the typical stellar populations in early-type
%spirals \citep{dejong96}.
%The \citet{holtzman95} prescription was used to convert the color excess
%to visual magnitudes of extinction \av.

IRAC images at 3.6  \micron\ %in all four bands (3.6, 4.5, 5.8, 8.0\,\micron)
were retrieved from the \spitzer\ archive. 
We started with the Basic Calibrated Data images,
and aligned and combined them with MOPEX \citep{makovoz05} which accounts
for distortion and rotates to a fiducial coordinate system.
1\farcs20 pixels were imposed for the final image, %all maps, 
roughly the same as the original IRAC detector.
Significant banding was present % in channels 1 and 2 (3.6, 4.5\,\micron)
from a bright star in the field, 
and this was corrected for by 
interpolation over the affected rows before combination with MOPEX.
%In channels 3 and 4 (5.8, 8.0\,\micron), banding was instead originating
%from \nnn's bright nucleus; the amplitude of the band was $\sim$0.45\,MJy\,sr$^{-1}$,
%or about 20\% of the background.
%These features were more difficult to correct for and we were forced to
%invent a slightly more sophisticated procedure applied to the
%individual images before combination. 
%We first calculated the column average of the
%band in two empty regions on opposite sides of the galaxy, and then fitted
%the one-dimensional (1D) mean of those with a Lorentzian. 
%This fit was finally reconverted into a 2D image and subtracted from the
%original.
Combination of these corrected images showed no discernible effect of
the original flaw.

The infrared and optical images will be discussed in Sections \ref{sec:stars}
and \ref{sec:dust}. 

\section{Molecular gas results \label{sec:coresults}}

\begin{figure*}[!th]
\centering
\includegraphics[width=\linewidth, bb=0 0 612 792]{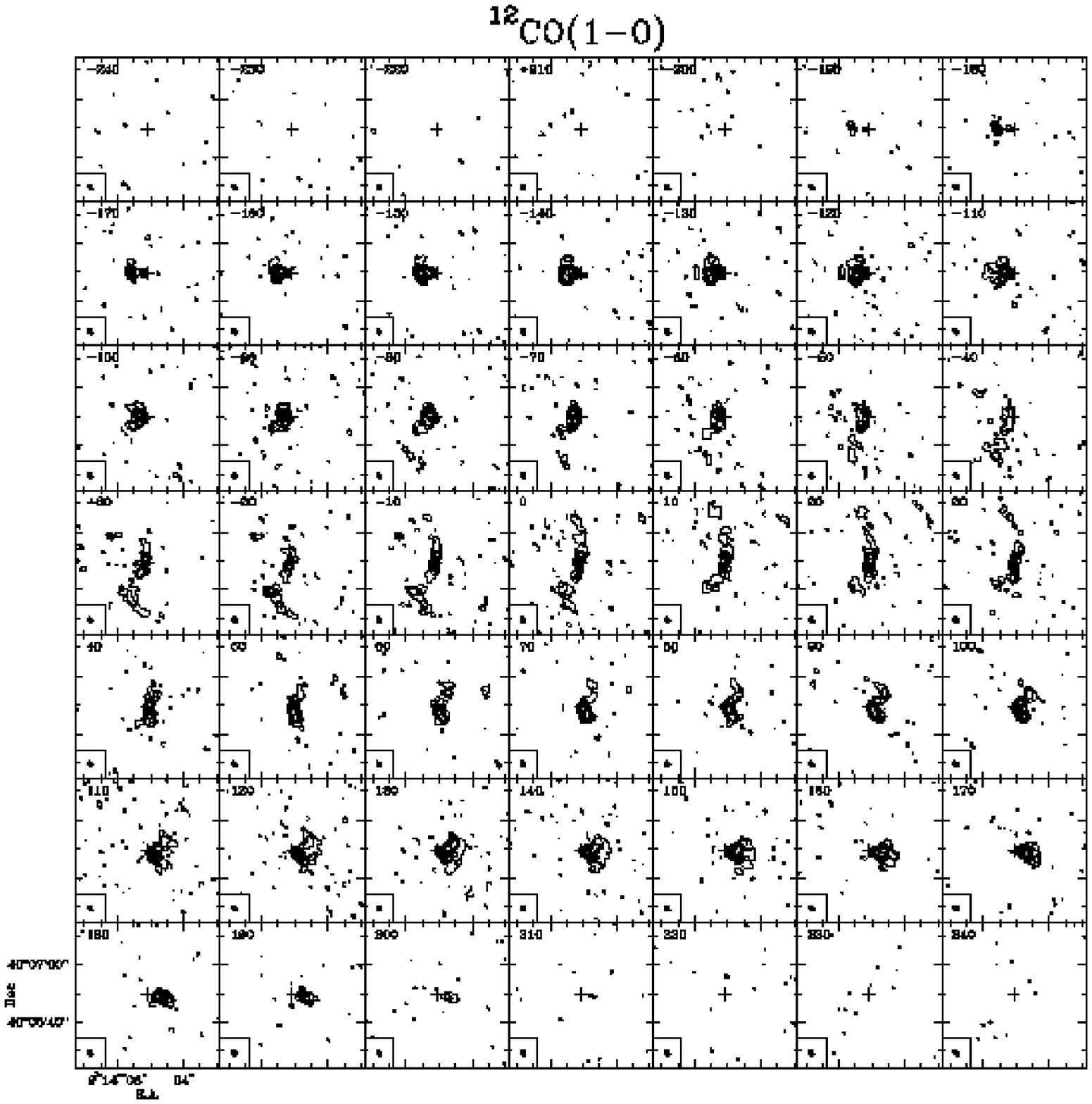}
\caption{ $^{12}$CO(1--0) velocity-channel maps observed with the
PdBI in the nucleus of NGC\,2782 with a spatial resolution of
2.1\arcsec$\times$1.5\arcsec\ at PA=38$^{\circ}$ (beam is plotted as a filled
ellipse in the bottom left corner of each panel). We show a field of view of
50\arcsec, i.e. $\sim$1.2 times the diameter of the primary beam at 115~GHz.
The phase tracking center is indicated by a cross at
$\alpha_{J2000}$=$09^h14^m05.08^s$ and $\delta_{J2000}$=$40^{\circ}06'49.4''$.
Velocity-channels are displayed from v=--240~km~s$^{-1}$ to v=240~km~s$^{-1}$
in steps of 10~km~s$^{-1}$. Velocities are in LSR scale and refer to
v=2562~km~s$^{-1}$. Contour levels are --3$\sigma$, 3$\sigma$,
7$\sigma$, 12$\sigma$, 19$\sigma$, 30$\sigma$ 40$\sigma$ and 55$\sigma$ where
the rms $\sigma$=2.0~mJy~beam$^{-1}$.  \label{fig:chan1} }
\end{figure*}

\begin{figure*}[!th]
\centering
\includegraphics[width=\linewidth, bb=0 0 612 792]{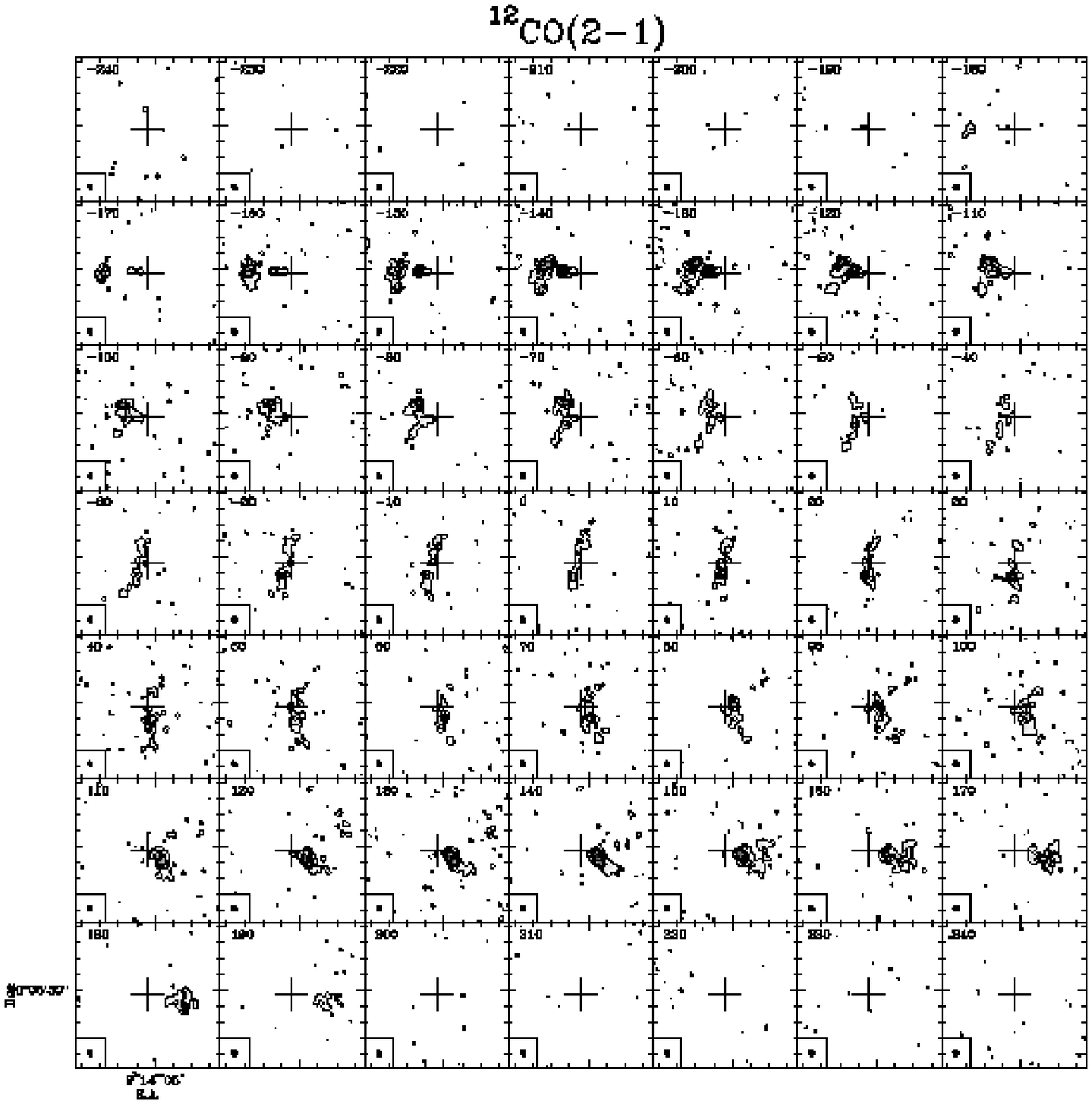}
\caption{Same as Fig.~\ref{fig:chan1} but for the 2--1 line of $^{12}$CO.
Spatial resolution reaches 0.7\arcsec$\times$0.6\arcsec\ at PA=20$^{\circ}$
(beam is plotted as a filled ellipse in the bottom left corner of each panel).
We show a field of view of 18\arcsec, i.e. $\sim$0.9 times the diameter of
the the primary beam at 230~GHz.  Velocity-channels are displayed from
v=--240~km~s$^{-1}$ to v=240~km~s$^{-1}$ in steps of 10~km~s$^{-1}$, with same
reference as used in Fig.~\ref{fig:chan1}. Contour levels are --3$\sigma$,
3$\sigma$, 7$\sigma$, 10$\sigma$, 14$\sigma$ and 18$\sigma$, where the 
rms $\sigma$=5.2~mJy~beam$^{-1}$. \label{fig:chan2}}
\end{figure*}

Figures \ref{fig:chan1} and \ref{fig:chan2} show the velocity-channel maps
of \coone\ and \cotwo\ emission in the central region of \nnn.
The kinematics show the typical signature of a rotating disk, 
together with non-circular motions more prominent to the south. 
The western (eastern) side of the CO bar is red- (blue-) shifted relative to
the systemic velocity \vsys, which is fitted to be at 2555$\pm$10\kms.
This agrees very well with the \hi\ heliocentric velocity of 2555 \kms \citep{smith94}.

The dynamical center was derived from the center of symmetry of the velocity field,
and chosen to maximize the velocity dispersion. 
It is determined to lie at ($\alpha_{2000}$, $\delta_{2000}$)\,=\,
(09$^{\rm h}$14$^{\rm m}$05.11$^{\rm s}$, 40$^\circ$06$^\prime$49.24\arcsec),
not far from the phase-tracking center. 
The position of the dynamical center coincides almost exactly with 
the compact radio source discovered by \citet{melanie07}:
($\alpha_{2000}$, $\delta_{2000}$)\,=\,
(09$^{\rm h}$14$^{\rm m}$05.11$^{\rm s}$, 40$^\circ$06$^\prime$49.32\arcsec). 
Both are also within the errors of the
position of the ``C'' component of the \nnn\ nucleus
described by \citet{saikia94}. 
Therefore we ascribe the position of the AGN and the dynamical center
of \nnn\ to the position derived here from the CO kinematics.

\subsection{CO morphology \label{sec:comorphology}}

\begin{figure}[!th]
\centering
\includegraphics[width=0.9\linewidth]{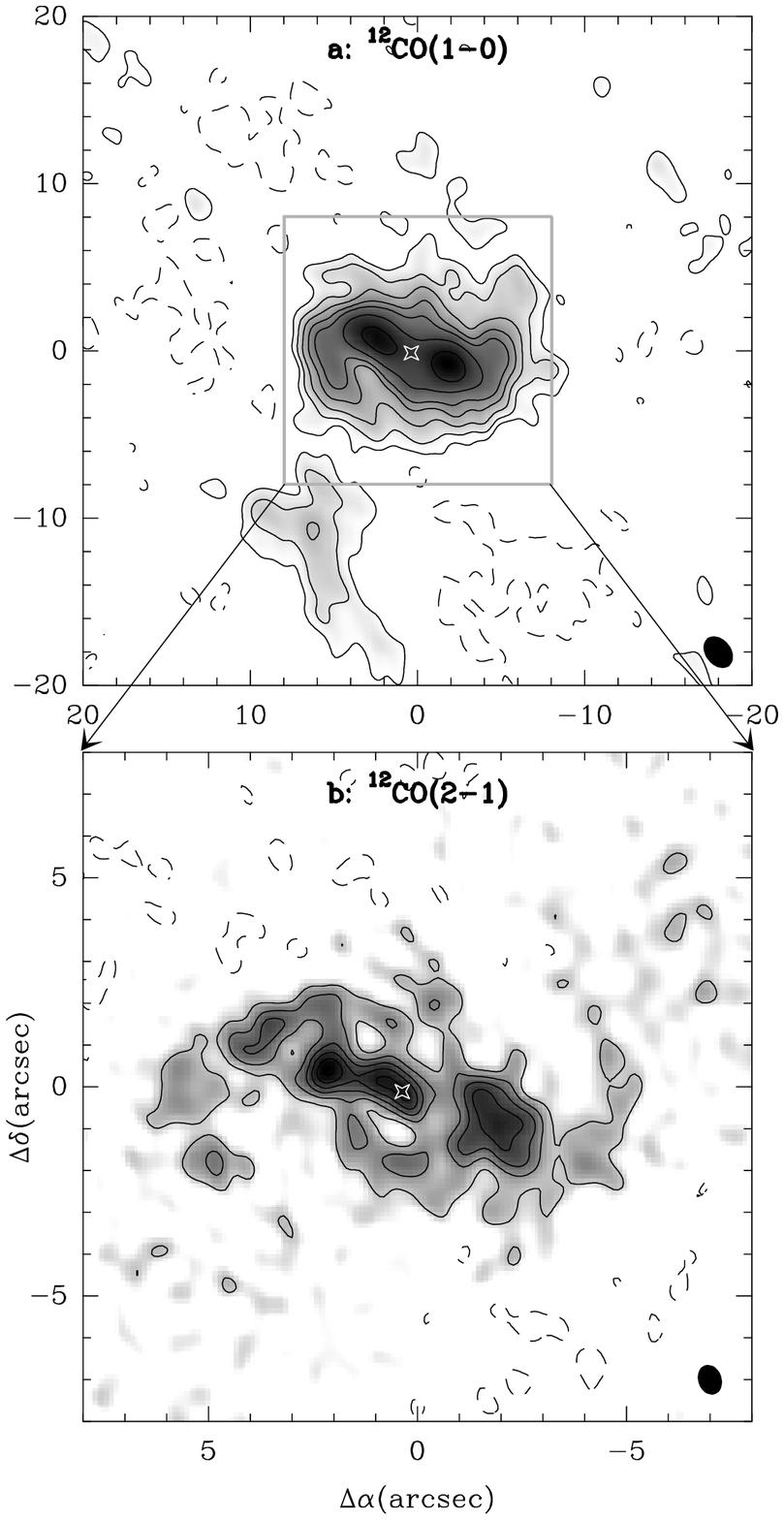}
\caption{Top panel: {\bf a}) The natural-weight map in \coone\ shown in grey scale
with contours ranging from 0.5 to 8.0 Jy\,\kms\ in 
%0.5 Jy\,km\,s$^{-1}$ intervals.
1 Jy\,km\,s$^{-1}$ intervals.
Bottom panel: {\bf b}) The uniform-weight map in \cotwo\ shown in grey scale with
contours ranging from 1 to 7.0 Jy\,km\,s$^{-1}$ in 
%0.5 Jy\,km\,s$^{-1}$ intervals.
1 Jy\,km\,s$^{-1}$ intervals.
\label{fig:cozero}}
\end{figure}

The overall distribution of the molecular gas is illustrated in
Figure \ref{fig:cozero} which shows the velocity-integrated CO intensity maps,
achieved by integrating channels from $v=-$230 to 230\,km\,s$^{-1}$. 
The naturally weighted \coone\ map is shown in the top panel of Fig. \ref{fig:cozero},
and the uniform-weight \cotwo\ map in the bottom panel.
After correcting for primary beam attenuation and considering only measurements
with a signal-to-noise of $3\sigma$ or greater on the data cube,
our naturally weighted (1--0) observations in the central 42\arcsec\ detect 
an integrated CO flux of 150 Jy\,km\,s$^{-1}$. 
This corresponds to 65\% of that in the 
Five College Radio Astronomy Observatory single dish measurements 
\citep[within 45\arcsec: 230 Jy\,km\,s$^{-1}$,][]{young95},
and roughly 78\% of the flux measured by \citet[][195 Jy\,km\,s$^{-1}$]{jogee99}.
Our measurements are sensitive to small-scale structure but are missing some
fraction of the diffuse component.
We use the short-spacing single dish observations by \citet{young95}
to derive the gas mass for numerical simulations in Sect. \ref{sec:code}. 

The \coone\ emission is distributed in a nuclear elongated structure 
already identified by \citet{jogee99}.
Our observations clearly delineate the
diffuse spiral arms extending to the north and south.
The spirals are not symmetric, and there is substantially more emission to the 
southeast than to the northwest.
The filaments at $\sim$10\arcsec\ to the north and south in CO(1--0) correspond to 
the spiral arms from the outer stellar oval (which we discuss in more detail
below).
Indeed, the CO ``spur'' labeled ``O2'' by \citet{jogee99} appears to be associated
with the extended southern spiral arm shown in our maps
(see Fig. \ref{fig:cozero}).

The central structure in the \cotwo\ map is clearly resolved;
the higher resolution afforded by our new observations
shows a clearly elongated structure with spiral arms commencing at the
ends of the elongation.
The gas in the inner spiral arms is aligned along the elongation with a pitch angle 
of $\sim90^\circ$, 
which makes the feature resemble more a bar than an inclined or edge-on disk. 
At the end of the feature, the gas changes direction to follow the outer spiral 
arms, which are situated at a lower pitch angle.
Moreover, the molecular gas is clearly responding to the stellar
oval/bar which we will discuss in Section \ref{sec:stars}.
Hence, we will refer to the circumnuclear molecular structure as
the ``nuclear gas bar'', as distinct from the nuclear stellar bar (see below).

The molecular gas mass within the 42\arcsec\ PdB primary beam field is
M$_{H_2}\,=\,1.4\times10^{9}$\,\msun, assuming the CO-to-\htwo\ conversion factor
$X\,=\,2.2\times10^{20}$\,cm$^{-2}$(K\,km\,s$^{-1}$)$^{-1}$ given by \citet{solomon91}.
Including the helium mass in the clouds (multiplying $\times1.36$) gives
M$_{{H_2}+He}\,=\,1.9\times10^{9}$\,\msun.
Most of the molecular gas mass in \nnn\ is in the nuclear gas bar$+$spiral, 
making it an extremely massive structure.
Indeed, the circumnuclear molecular gas component in \nnn\ is particularly massive,
roughly 3 times more so than most of the NUGA galaxies studied so far
(NGC\,4826: \citealt{santi03}, NGC\,7217: \citealt{francoise04}, NGC\,3718: \citealt{melanie05}
NGC\,4579: \citealt{santi05}, NGC\,6951: 
\citealt{santi05} all have molecular masses on the order of $\sim3\times10^8$\,\msun).
Only NGC\,4569 \citep{fred07}, with M$_{H_2}\,=\,1.1\times10^{9}$\,\msun, is
roughly comparable with \nnn.

\subsection{Kinematics \label{sec:kinematics}}

\begin{figure}[!th]
\centering
\includegraphics[width=\linewidth, bb=150 100 483 680]{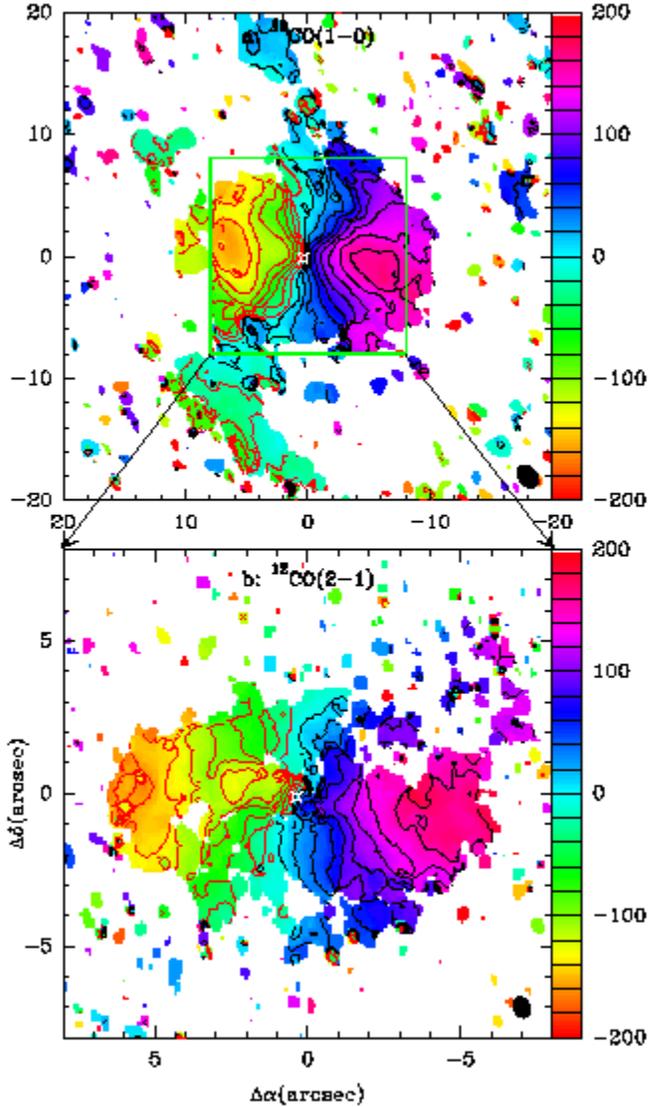}
\caption{ $^{12}$CO(1--0) (a: top panel) and  $^{12}$CO(2--1) (b: bottom)
isovelocities contoured over false-color velocity maps.
The AGN (dynamical center) position is marked with a star.
\label{fig:isovels}
}
\end{figure}

Figure \ref{fig:isovels} shows the mean velocity field derived from
CO(1--0) (top panel) and CO(2--1) (bottom).
The star marks the position of the AGN (coincident with the dynamical
center).
The kinematic signature of a rotating disk is clearly seen in Fig. \ref{fig:isovels},
and in the body of the circumnuclear molecular spiral, there are
few non-circular motions.
However, the southern spiral arm in the CO(1--0) map (top panel of Fig. \ref{fig:isovels})
shows clear streaming motions, and
appears to be somewhat decoupled from the nuclear spiral because of
the velocity discontinuity toward the southeast.
In the bottom panel of Fig. \ref{fig:isovels},
systematic kinks appear in the CO(2--1) velocity field 
near the position of the AGN ($\sim$0\farcs5, barely resolved) both to the north and the south.

We derive a kinematic major axis of PA = (75\,$\pm$\,5)$^\circ$, 
consistent with previous determinations \citep[75$^\circ$,][]{jogee99}.
We will assume that the inclination angle of the nuclear spiral in \nnn\ is 
$\sim$30$^\circ$ \citep[see also][]{jogee99}.
A smaller inclination would result in $\sin i$ corrected velocities which would be
too large to be consistent with even the largest spiral galaxies and ellipticals.
A larger inclination seems equally unlikely because of the relatively straight 
extended spiral arms (see discussion in Sect. \ref{sec:dust});
if the inclination were $\ga$30$^\circ$, the arms would be 
apparently more compressed relative to the line of nodes.
That the galaxy disk is inclined at 30$^\circ$ is confirmed by the
two-dimensional bulge-disk decomposition and the elliptical isophote fitting
described in Sect. \ref{sec:stars}.

\begin{figure}[!ht]
\centering
\includegraphics[width=0.9\linewidth]{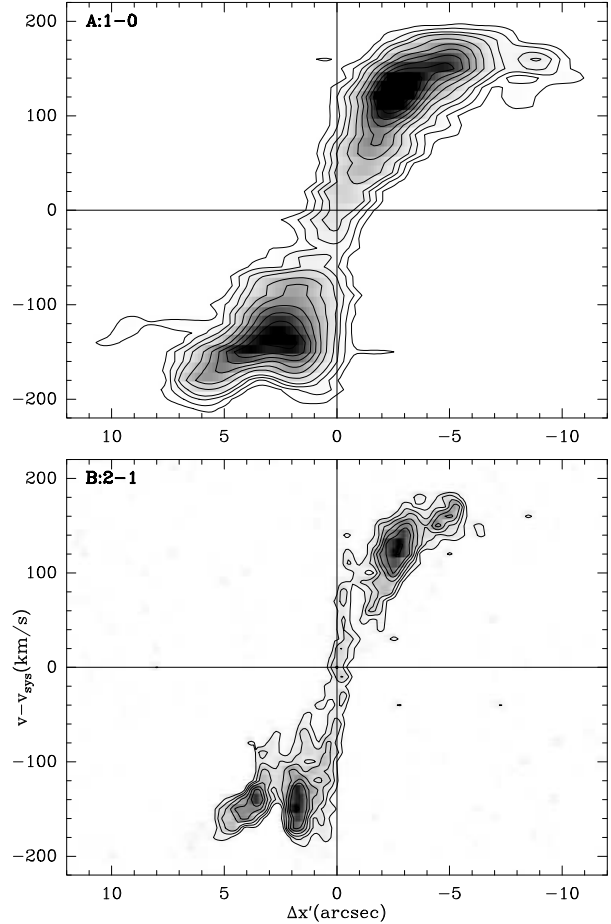}
\caption{Top panel: {\bf a}) Position-velocity diagram of \coone\ along the kinematic major axis
of \nnn\ (PA\,=\,75$^\circ$) contoured over a grey-scale representation.
Velocities have been rescaled to \vsys\,=\,2555\kms, and offsets are relative
to the dynamical center.
Bottom panel: {\bf b}) The same for \cotwo.  
\label{fig:pvmajor}}
\end{figure}

\begin{figure}[!ht]
\centering
\includegraphics[width=0.9\linewidth]{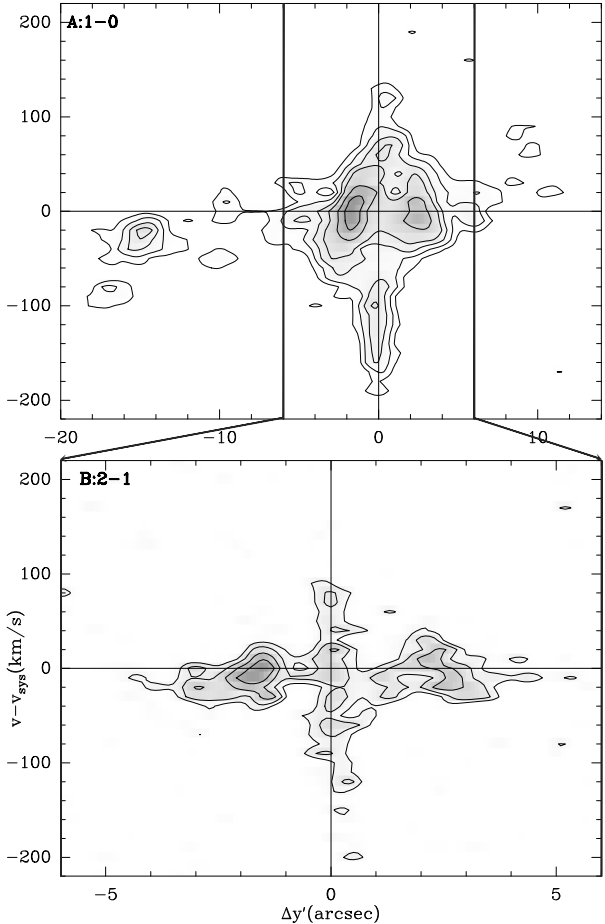}
\caption{Top panel: {\bf a}) Position-velocity diagram of \coone\ along the kinematic minor axis
of \nnn\ (PA\,=\,165$^\circ$) contoured over a grey-scale representation.
Bottom panel: {\bf b}) The same for \cotwo.  
\label{fig:pvminor}}
\end{figure}

Position-velocity (PV) diagrams along the major kinematic axis of \nnn\
are shown in Fig. \ref{fig:pvmajor}, and along the minor axis in Fig. \ref{fig:pvminor}.
In both figures, \coone\ is given in the top panel, and \cotwo\ in the bottom.
The top panel of Fig. \ref{fig:pvmajor} reveals regular circular rotation together
with mild streaming motions in the molecular gas.  
However, the rough alignment of the molecular feature at PA$\sim$88$^\circ$
with the line-of-nodes of the galaxy (PA$\sim72-73^\circ$, see Sect. \ref{sec:stars})
inhibits a clear signature of non-circular streaming motions.
Consequently,
the major-axis PV diagram cannot provide
conclusive evidence for out-of-plane kinematics or for kinematic decoupling of the 
nuclear structure from the larger-scale stellar disk.
The kinematics shown by the minor-axis PV in Fig. \ref{fig:pvminor}
are quite regular in the extended regions, 
but show significant velocity dispersion close to the nucleus.
This could reflect an unresolved rotational velocity component,
suggesting a large central dynamical mass (see below).
Finally,
Fig. \ref{fig:pvminor} shows no evidence for molecular gas outflow, as tentatively
suggested by \citet{jogee99}.

\subsection{The rotation curve and dynamical mass \label{sec:rc}}

We have derived a rotation curve (RC) from the PV diagram taken along
the kinematic major axis of \nnn\ at a PA = 75\,$^\circ$. 
The terminal velocities were derived by fitting multiple Gaussian
profiles to the spectra across the major axis. 
The fitted velocity centroids, corrected for $\sin i$ ($i\,=\,30^\circ$),
give \vrot\ for each galactocentric distance. 
Although for both lines the RCs to the south 
(negative velocities) are slightly steeper within the inner 300\,pc than
those to the north (positive), we averaged together the two
curves derived from either side of the major axis. 
Hence, the resulting \vrot\ may be slightly shallower in the
inner regions than the true mass distribution would imply.
Because data for both lines were consistent, we combined both curves
into an average by spline interpolation.
The final RC is shown in Sect. \ref{sec:interpretation} in the context of
our numerical simulations.

The peak velocity is $\sim170/\sin\,i$ \kms, or $\sim$340\,\kms\ for
$i\,=\,30^\circ$, obtained at a radius of 850\,pc (5\arcsec) to the south,
but at $\sim$1\,kpc ($\sim$6\arcsec) in the RC obtained from averaging
both lines and both sides of the major axis.
%According to the precepts set out by \citet[][see also \citealt{santi03}]{lequeux83}, 
From the peak velocity, we can derive a rough estimate of the
dynamical mass: $M(R)\,=\,2.32 \times 10^5 R\, V^2(R)$
where $M(R)$ is in \msun, $R$ in kpc, and $V$ in \kms.
Assuming the most flattened disk-like distribution (i.e., including a multiplicative
constant of 0.6), at a radius of 1\,kpc,
we would find $M_{dyn}\,=\,1.6\times10^{10}$\msun.
Within a 1.7\,kpc radius, where virtually all of the observed molecular gas is located,
assuming a flat rotation curve (see Sect. \ref{sec:interpretation}), we would
infer $M_{dyn}\,=\,2.7\times10^{10}$\msun. 
Comparing with the mass in molecular gas including helium would give
a molecular mass fraction of $\sim$7\% in this region,
in agreement with \citet{jogee99}.

%Along the minor axis PV, the high resolution of our CO(2--1) observations
%enables an estimate of the mass within a radius of $\sim$100\,pc (0\farcs6).
%With a velocity of 110$\sin i$ \kms, we find a dynamical mass of $\sim10^{9}$\msun;
%$\sim$4\% of the total dynamical mass within a $\sim$3.4\,kpc diameter is
%concentrated in a region of diameter $\sim$200\,pc.

\section{Stellar structure \label{sec:stars}}

\begin{figure}[!ht]
\centering
\includegraphics[width=\linewidth, bb=150 100 483 680]{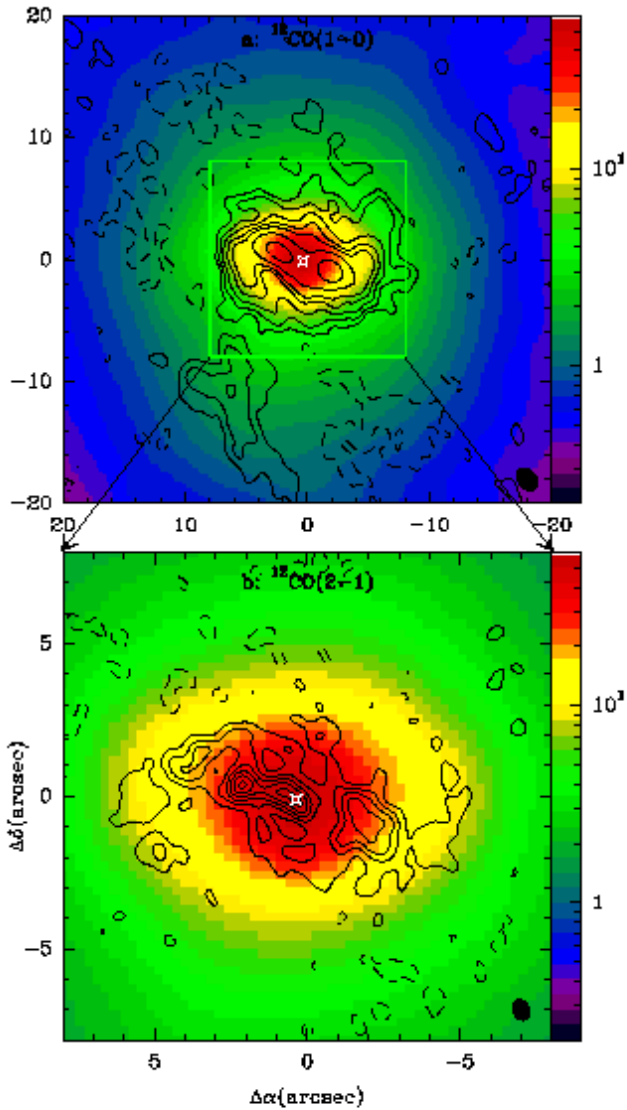}
\caption{Top panel: {\bf a}) \coone\ total intensity contoured over a false-color
representation of the IRAC 3.6\micron\ image of \nnn.
The nuclear bar/oval and the outer oval can clearly be seen at
PA$\sim$88$^\circ$ and $\sim$10$^\circ$, respectively. 
Bottom panel: {\bf b}) The same for \cotwo.  
In both panels, the AGN (dynamical center) position is marked with a star.
\label{fig:coirac1}}
\end{figure}

Figure \ref{fig:coirac1} shows the CO(1--0) and (2--1) total intensity
maps overlaid on the IRAC 3.6\micron\ image described in Sect. \ref{sec:otherdata}.
This wavelength traces very well the massive component of the stellar
populations in galaxies, and has the added advantage of very low extinction
even compared to the $K$ band.
Two main inner structures can be seen in Fig. \ref{fig:coirac1}: 
an {\it inner stellar bar/oval} with
a PA$\sim$88$^\circ$ and diameter of $\sim$15\arcsec,
and an {\it outer oval} at PA$\sim$10$^\circ$ and diameter of $\sim$30\arcsec\footnote{These
values have been derived by fitting elliptical isophotes to the image with
the task {\it ellipse} in the IRAF/STSDAS package.}.
Previous $K$-band observations \citep{jogee99} identified 
similar features, but with slightly different PA's and
a larger radius for the outer oval than we find here
(25\arcsec\ vs. $\sim$15\arcsec). 
Both structures are clearly present in \nnn,
and as we shall see in Sect. \ref{sec:torques}, contribute to the
dynamical perturbations in this galaxy.

\begin{figure*}[!ht]
\centering
\hbox{
\includegraphics[width=0.35\linewidth, bb=-14 -63 629 857, angle=-90]{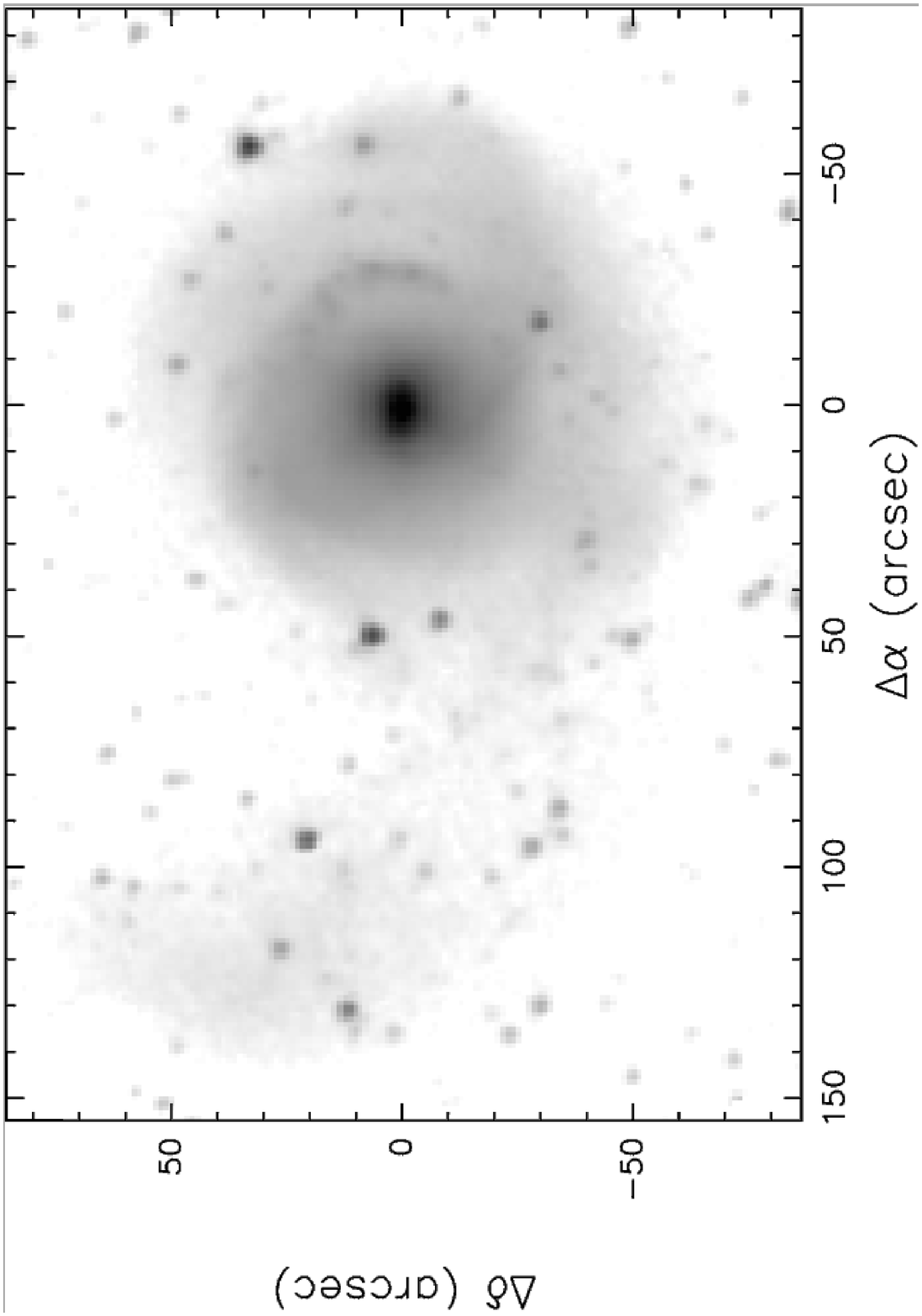}
\includegraphics[width=0.35\linewidth, bb=-14 -52 629 846, angle=-90]{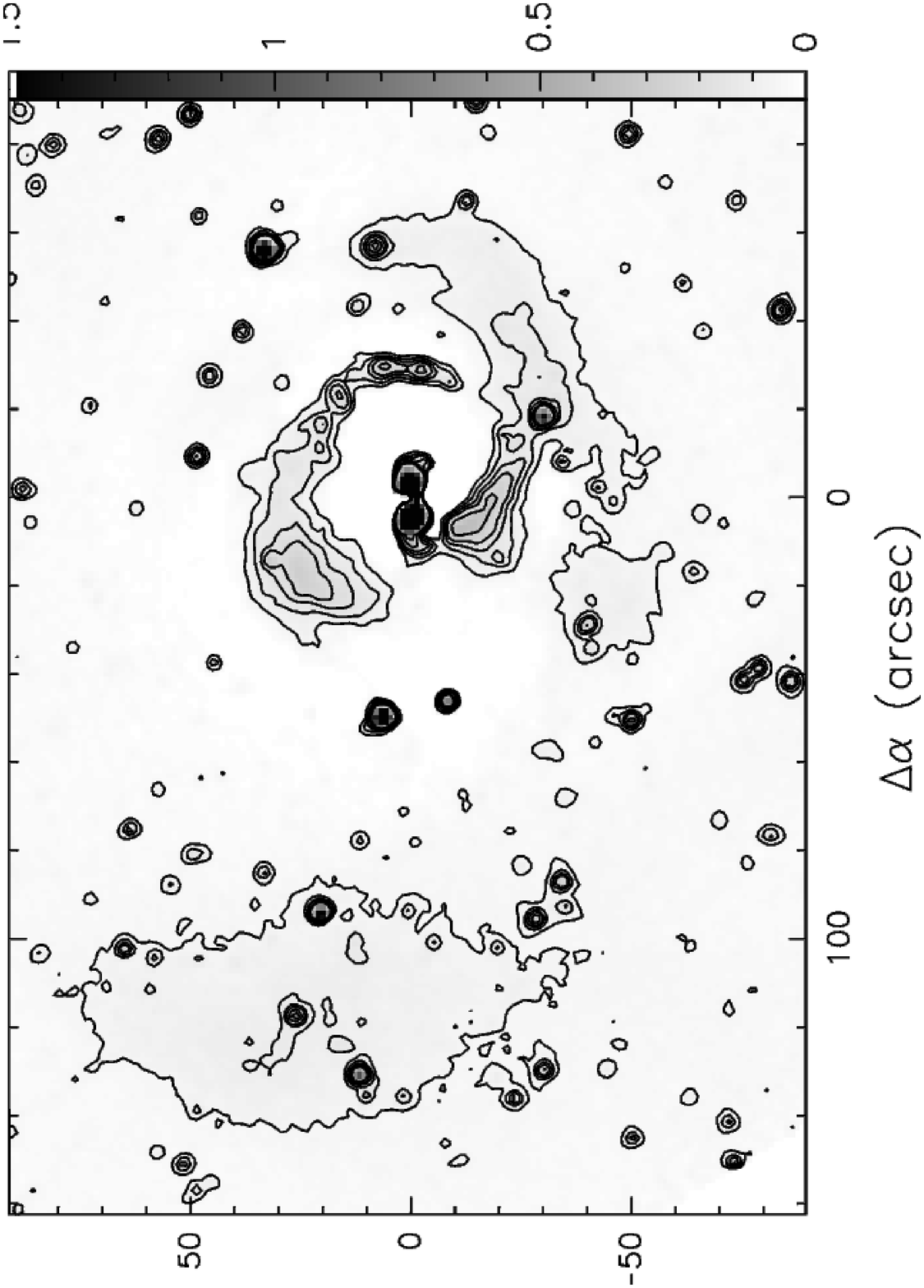}
}
\caption{{Left panel: \bf a}) IRAC 3.6\,\micron\ image of \nnn. 
The stellar sheets to the east and west, and the ripples about $\sim$40\arcsec\
to the east are clearly visible.
Right panel: {\bf b}) IRAC 3.6\,\micron\ residuals from the bulge-to-disk
decomposition described in Sect. \ref{sec:bd},
with contours superimposed.
\label{fig:irac1}}
\end{figure*}

A large-scale view of \nnn\ at 3.6\,\micron\ is illustrated in 
(the left panel of) Fig. \ref{fig:irac1}.
The detached stellar sheet or tail outside the main galaxy disk is evident
to the east, and the elongation or distortion of the disk to the west is clearly visible.
There are also stellar arcs or ``ripples'' \citep{smith94,jogee99} about 25\arcsec\
to the west of the nucleus.
All these features are seen in optical images \citep{smith94,jogee99}
and in \hi\ \citep{smith91}.
Moreover, in \hi, there is a long ($\sim$\,54\,kpc) plume or
tidal tail extending to the north, 
with its origin at the westernmost edge of the distorted disk \citep{smith91}.

\subsection{Bulge/disk decomposition \label{sec:bd}}

To better understand the mass distribution and investigate the non-axisymmetric
structure in the stars,
we performed a two-dimensional bulge/disk decomposition on the IRAC 3.6\micron\
image of \nnn\ with {\it galfit}, the publicly available algorithm developed by
\citet{peng02}.
The MOPEX point-response function was used for convolution with the image, and
the background sky level was fixed to the measured value rather than fit.
Initial parameters for bulge and disk were guessed by using the scaling relations 
given by \citet{moriondo98} and \citet{hunt04}.
The final best fit was achieved with a generalized exponential (Sersic) bulge,
an exponential disk, and a nuclear point source.

We ran several sets of fits, in order to experiment with masking the stellar 
tails or sheets predominant to the east.
Fitting the whole image ($\sim$3.4\arcmin diameter) results in a bulge with
shape index $n\,=\,3$ (a de Vaucouleurs bulge has $n\,=\,4$), 
an effective radius of 1.1\,kpc and an apparent inclination of 32$^\circ$.
This bulge contains about 60\% of the total 3.6\micron\ light. 
The resulting disk, with a scalelength of 2.9\,kpc and an inclination of 41$^\circ$, 
together with a nuclear point source
contribute about 33\% and 7\%, respectively, to the total luminosity.
The fitted aspect ratio of the disk is very close to that given for \nnn\ in NED, which
corresponds to a system inclination of 42$^\circ$. 
The bulge PA of 19$^\circ$ is ill-determined because of its low apparent flattening,
but the fitted orientation of the disk (PA\,=\,73$^\circ$) is consistent with
the angle of the line-of-nodes estimated by \citet{jogee99}
and with the kinematical major axis found in Sect.\ref{sec:kinematics}.

Masking the stellar ``sheets'', and confining the fit to the undisturbed portion of the
outer regions (within a $\sim$2.4\arcmin diameter, 24\,kpc) gives a slightly different,
probably more reliable, fit.
The bulge has a steeper shape index ($n\,=\,4$), the nuclear point source is smaller, 
and the disk is less inclined.
The fraction of bulge-to-total luminosity remains the same, $\sim$60\%.
The masked fits give a disk inclination of 33$^\circ$, more similar to that of the bulge,
and more consistent with the visual aspect of the galaxy. 
Indeed, inspection of Fig. \ref{fig:irac1} suggests a strikingly round
appearance, at least in the regular portion of the disk out to a radius of $\ga$50\arcsec\ 
($\sim$9\,kpc).
This impression is confirmed by fitting the isophotes to ellipses\footnote{With the 
task {\it ellipse} in the IRAF/STSDAS package.} which
shows that at galactocentric distances as large as 12.5\,kpc, the 
fitted system ellipticity implies an inclination of $\la33^\circ$ at PA\,=\,72$^\circ$.
Hence, we confirm $\sim$30$^\circ$ for the system inclination (as described in
Sect. \ref{sec:kinematics}), obtained with the masked bulge-disk decomposition. 

\begin{figure*}[!th]
\includegraphics[width=1.0\linewidth]{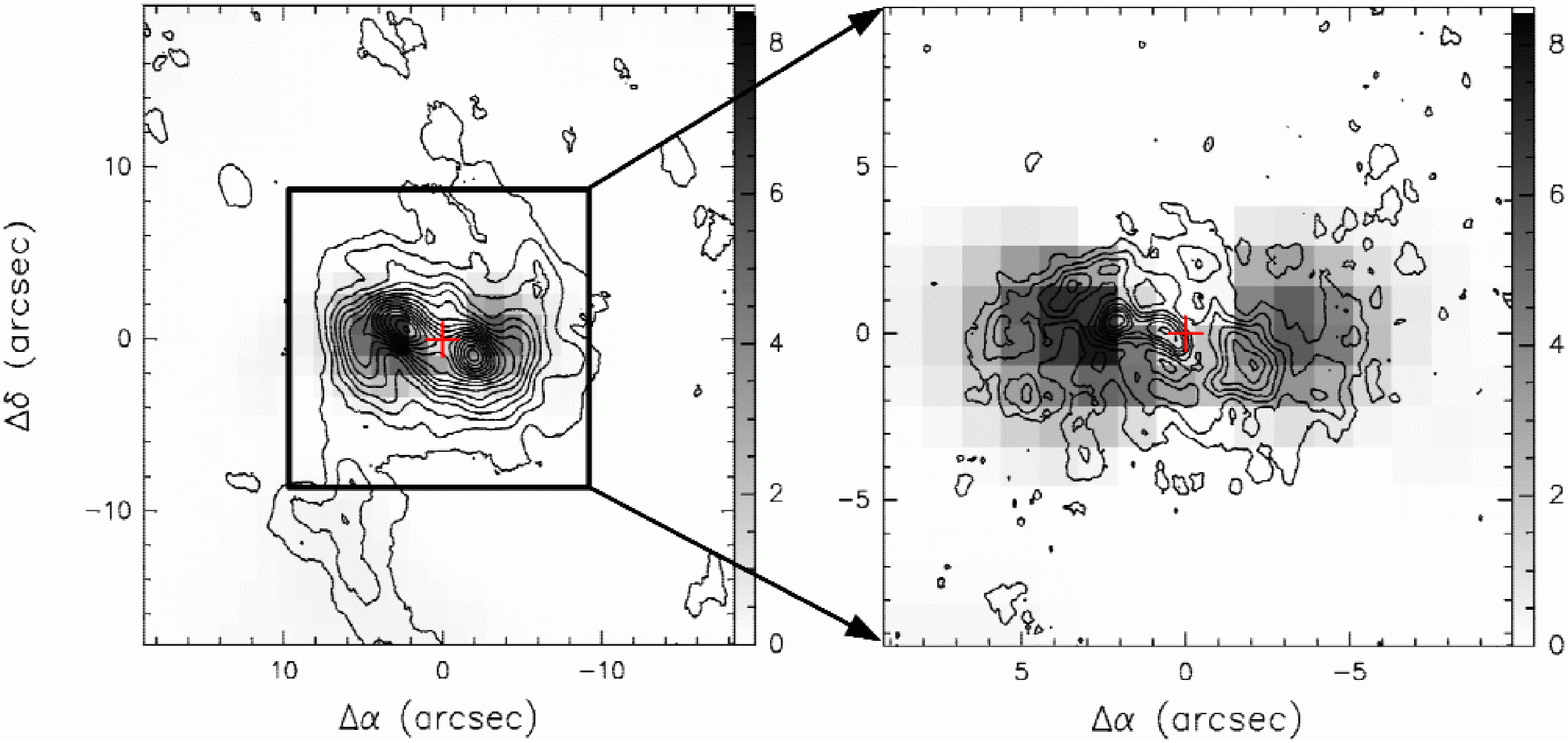} 
\caption{Left panel: {\bf a)} shows CO(1-0) total intensity contours overlaid on the
gray-scale {\it residuals} (in MJy/sr)
of the bulge-disk decomposition performed on the 3.6\micron\ IRAC image. 
Right panel: {\bf b)} shows the analogous overlay of CO(2-1).
The phase-tracking center is shown by a cross.
\label{fig:bdres_co} }
\end{figure*}

The fit provides a convenient axisymmetric model for unsharp masking.
The large-scale residuals from the fit described above are shown in the right panel of Fig. \ref{fig:irac1}.
The correspondence with the features in the unsharp masked optical images \citep{smith94}
is excellent.
The stellar sheet to the east, and the ripples and distortion of the disk to the west,
are clearly revealed.
The small-scale residuals of the bulge/disk decomposition are shown in Fig. \ref{fig:bdres_co}
with CO(1-0) (left panel) and CO(2--1) (right) overlaid in contours;
both the ``masked'' fits and the fits to the entire image give virtually
identical residuals in the circumnuclear region.
The large (1\farcs2) pixels of the IRAC image impede detailed comparison, but
the residuals from the fit have an $m=2$ structure, and strongly resemble a
stellar bar.
Some small component of residuals could perhaps be hot dust rather than stars,
because of the red $K-L$ color excess observed in the center of \nnn\ \citep{hunt92}. 
Nevertheless, dust cannot contribute significantly to the structure of the residuals
because the $K-L$ color is redder outside the bar-like structure than within it.
In fact,
the position of the bar-like residuals coincides perfectly with the CO emission,
and provides a classic illustration of the theoretical behavior of the 
spiral response of a gas component to bar forcing.
In this case, the gas is phase shifted in advance relative to the bar (leading), 
as can be seen to the northeast and the southwest in the ``S'' shape of the spiral. 

\section{Star formation and dust in \nnn \label{sec:dust}}

\begin{figure}[!t]
\centering
\includegraphics[width=0.9\linewidth]{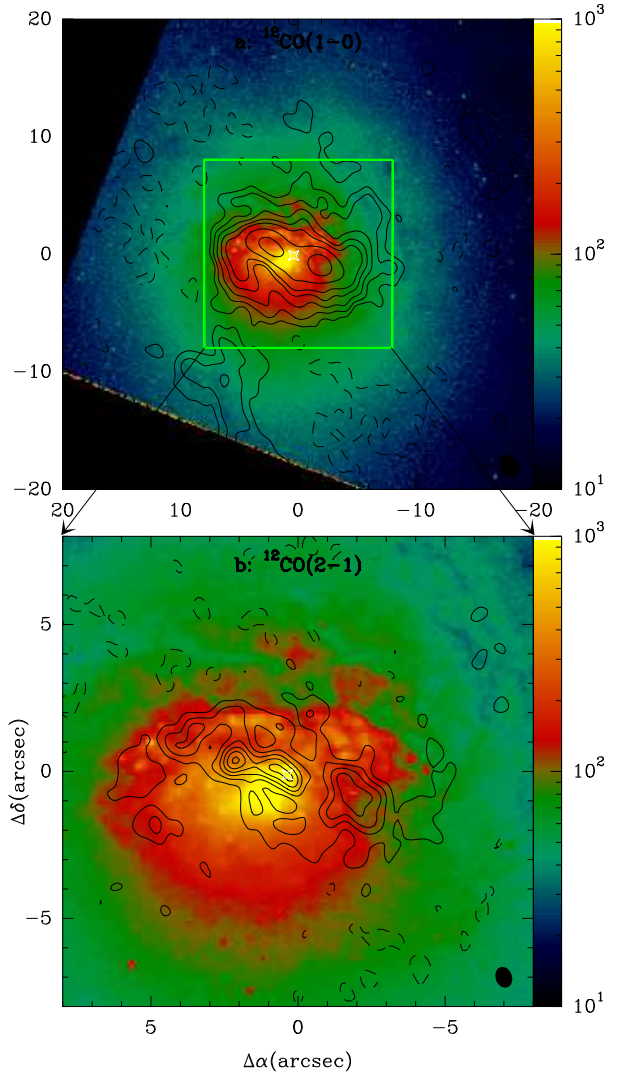}
\caption{Top panel: {\bf a}) \coone\ total intensity contoured over a false-color
representation of the HST/WFPC2 F814W image of \nnn.
Bottom panel: {\bf b}) The same for \cotwo.  
In both panels, the AGN (dynamical center) position is marked with a star.
\label{fig:cohsti}}
\end{figure}

While stellar structure can be more readily inferred from near-infrared (NIR)
wavelengths, dust extinction and star formation are best investigated
in the optical.
Figure \ref{fig:cohsti} shows the CO intensity maps overlaid on the HST/WFPC2
F814W image.
The F814W emission of the circumnuclear region shows a strong asymmetric excess to the
northwest of the AGN (hereafter called the ``NW excess''), roughly coincident with the 
structure in the \ha\ maps \citep{jogee99}.
This excess is contiguous with an entire arc of bright knots to the north running from east 
to west, again reflecting the \ha\ emission-line morphology.
Part of this arc lies within the \cotwo\ emission, while the excess to the
northwest is outside of it.
There is also an extended F814W excess generally to the south,
leading to a lopsided appearance at this wavelength.

%\begin{figure}[!ht]

\begin{figure}
\centering
\includegraphics[width=\linewidth, bb=150 100 483 680]{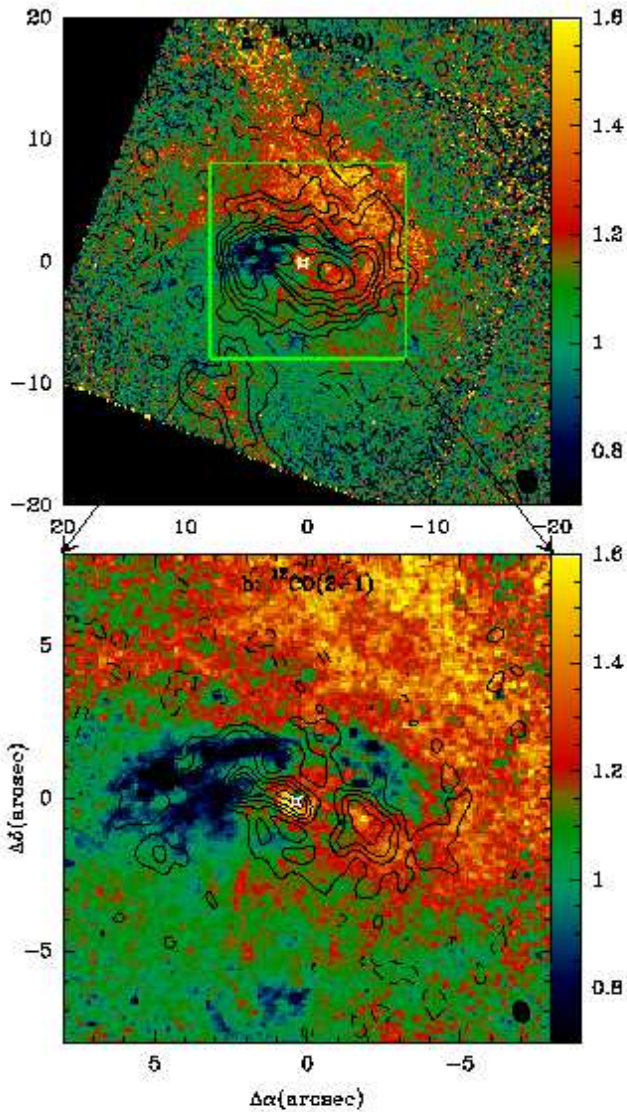}
\caption{Top panel: {\bf a}) \coone\ total intensity contoured over a false-color
representation of the HST/WFPC2 \vi\ image of \nnn.
The spiral arm to the northwest is clearly traced by the CO(1--0) emission.
Bottom panel: {\bf b}) The same for \cotwo.  
In both panels, the AGN (dynamical center) position is marked with a star.
\label{fig:cohstvi}}
\end{figure}

Figure \ref{fig:cohstvi} presents the CO intensity maps overlaid on the HST/WFPC2
\vi\ image described in Sect. \ref{sec:otherdata}.
The red \vi\ color delineating the spiral arm to the northwest is neatly traced
by the CO(1--0) emission.
This red arm connects to the CO spiral with a hook-like structure which winds
around the NW excess, following the CO emission to the nucleus.
The blue \vi\ colors of the F814W arc to the north of the AGN 
suggest that it is tracing the same star-formation episode revealed by \ha.
%as already implied by the similarity of the morphology.

%\begin{figure}[!ht]

This northern arc of star formation was identified by \citet{jogee98} and \citet{jogee99},
and occurs roughly cospatially with an outflow thought to be driven by the starburst. 
The outflow is seen at radio wavelengths and in optical emission lines
\citep{boer92,saikia94,jogee98,yoshida99,jogee99}. 
It is relatively north-south symmetric in the 6\,cm radio continuum \citep{saikia94},
but has the form of a ``mushroom'' in the optical
\citep[clearly illustrated by Fig. 8 in ][]{jogee99}.
The arc defines the ``cap'' of the mushroom, and the slightly narrower
``stem'' extends southward, ending at the blue knot roughly $\sim$7\arcsec\
south and 1\arcsec\ east of the nucleus.
This mushroom shape is not clearly seen in the \vi\ image.
The roughly normal \vi\ colors ($\sim$0.8--1) to the south of the CO(2--1) emission 
suggest that the unreddened stellar population of the bulge is obscuring
the star formation to the south,
a conclusion which we will motivate in the following.

The three-dimensional orientation of the galaxy disk can be inferred on the basis
of the red \vi\ colors in the extended spiral arm in the northwest.
The roughly symmetric spiral arm to the southeast is not reddened,
implying that the spiral structure to the northwest is being seen through an obscuring
dust lane, coplanar with the inclined stellar disk. 
If this is true, then the north would be the near side of the disk, and the
south would be the far side. 
This orientation would also mean that the spiral arms are trailing, which
is the most probable configuration.
The implication is that, toward the south, the stellar disk is seen through the 
intervening bulge. 
The relatively small bulge effective scalelength of 5\arcsec$-$6\arcsec\ means that 
$\sim$50\% of the bulge light lies within the region of the CO emission.
To the south,
the bulge stars with normal \vi\ colors are overwhelming the young blue stars
in the star-formation event. 
Hence, the southern part of the ``mushroom'' structure is not observed in \vi,
but in emission lines, 
since the bulge is not expected to contain a significant amount of dust, 
nor does it emit contaminating line emission. 
To the south most of the bulge is in front of the intervening dust lane, 
while to the north, dust obscures more of the bulge
because most of the bulge light is behind it.
This interpretation would also help explain the 
F814W image with its apparent lopsidedness to the south.
In any case, the dust must be clumpy, and 
inhomogeneities in the dust distribution 
allow random regions behind the dust to shine through.

The AGN (dynamical center)
is located at the northwest border of a tiny red arc $\sim$80\,pc to the southeast 
(``parachute''-like, see Fig. \ref{fig:cohstvi}).
In this region, there are broad \ha\ lines which show considerable north-south asymmetry,
unlike the narrow component of the emission lines which is
relatively symmetric \citep{yoshida99}.
Interestingly, this red arc in the color map is at the same position as the
velocity discontinuities in the CO(2--1) spider diagrams (see Sect. \ref{sec:kinematics}).
Because of the feature's extremely small size, we can not make
any reliable statements, but speculate that it
could be the color (and kinematic) signature of a small-scale AGN outflow out of the 
plane of the disk. 

\section{Gravitational torques on the molecular gas \label{sec:torques}}

Streaming motions along the bar led \citet{jogee99} to conclude that there could be 
gas inflow feeding the starburst.
However, streaming motions by themselves do not imply any transfer of
angular momentum, nor can they be used to infer the direction of such transfer. 
Hence, 
we now examine the torques exerted by the stellar potential on the molecular gas
in \nnn, in order to assess whether the gas is, on average, flowing in or out.
Although our PdBI map recovers $\sim$65\% of that from single-dish measurements,
in the following analysis what matters is the structured gas component.
The diffuse large-scale molecular gas contributes equal quantities of
positive and negative torques on the gas orbits; hence it does not
influence the net torque.

There is no clear evidence for non-coplanarity of the CO nuclear bar$+$spiral and the  
large-scale stellar disk.
The stellar bar is oriented at PA$\sim88^\circ$, and the kinematic major axis
at $\sim75^\circ$, somewhat hampering the identification of non-circular
motions. 
Nevertheless, were the two structures to be non-coplanar,
we would expect to see some kinematic evidence which is not seen.
Hence, in disagreement with the schematic model presented by \citet{yoshida99}, 
in what follows, we
assume that all the flattened structures (nuclear CO spiral, nuclear stellar bar,
and outer stellar oval) lie within the same disk plane.

To derive the torques exerted by the stars on the gas,
we adopt the method presented in \citet{santi05} which relies on a stellar potential
$\Phi$ derived from a NIR image and on the gas response as inferred from the
CO maps.
We assume that the total mass is determined by the stars,
essentially invoking an extreme ``maximum disk'' solution \citep[e.g.,][]{kent87},
in which the gas self-gravity and the dark matter halo are neglected.
We further assume that the measured gas column density $N(x,y)$ derived from the CO
total intensity maps is a reliable estimate of the probability of finding gas at $(x,y)$
in the current epoch.
This is reasonable, namely that CO traces the total gas content, 
since atomic gas in the nuclei
of galaxies is typically a very small fraction of the total gas mass.

As emphasized in \citet{santi05}, a single key assumption underlies the 
validity of our estimate of angular momentum transfer in the gas:
we implicitly assume that the gas response to the stellar potential is roughly
stationary in the reference frame of the potential over a few rotation periods.
This would not be the case 
in the presence of strong self-gravity in the gaseous components.
However, otherwise, even in galaxies with several stellar pattern speeds at different 
radii in the disk, numerical simulations suggest that the gas response adjusts its
response to the dominant stellar configuration at a given radius
\citep[see][and references therein]{santi05}.

\subsection{The stellar potential \label{sec:potential}}

\begin{figure*}
\centering
% something weird about the old version of this image, 
% needs to be diminished in size, bb doesn't seem to work
%\includegraphics[width=0.5\linewidth,angle=-90]{n2782NEW-dep-pot-co.eps}
\includegraphics[width=0.5\linewidth,angle=-90, bb=160 80 510 742]{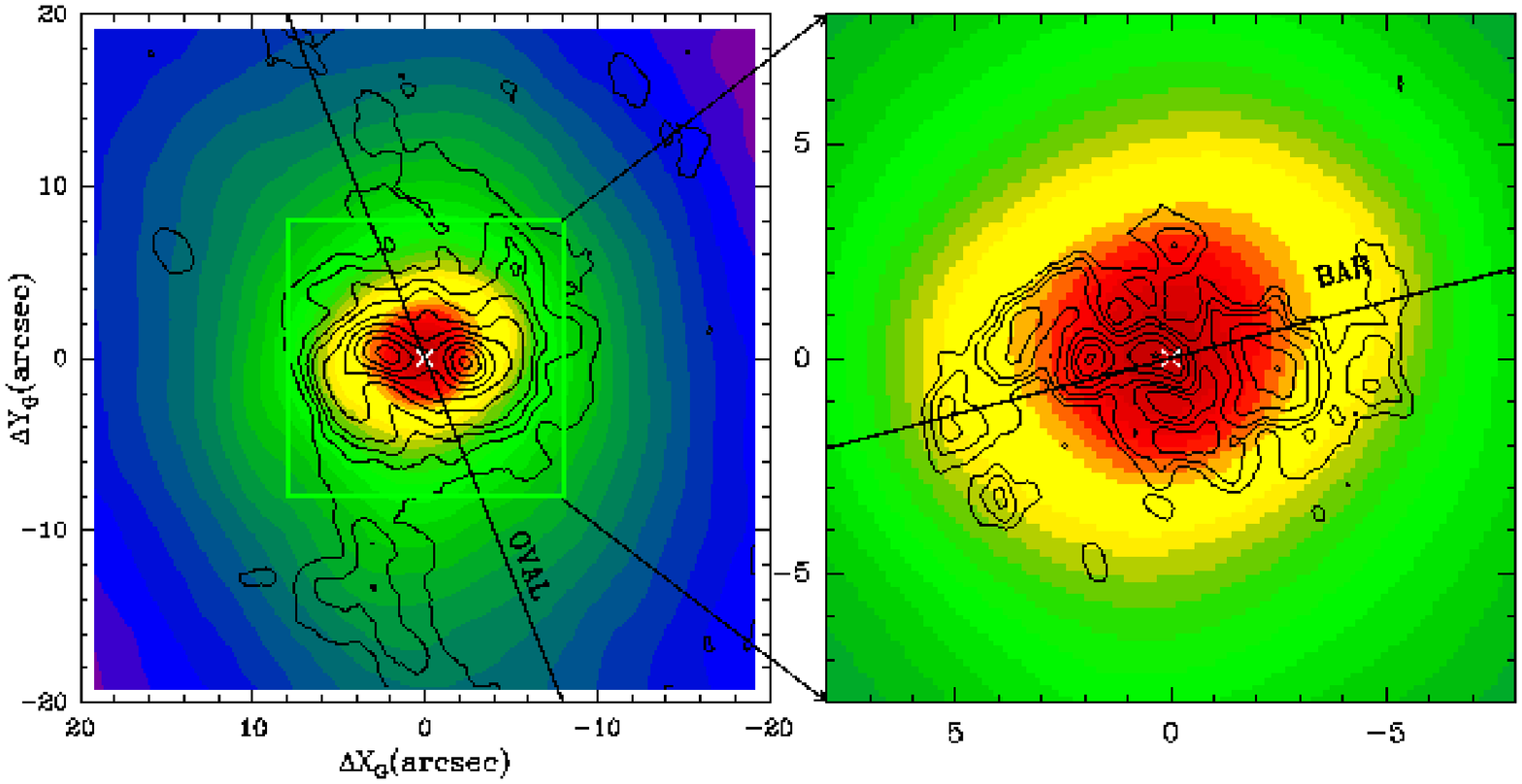}
\caption{Left panel: {\bf a}) \coone\ total intensity contoured over a false-color
representation the IRAC 3.6\micron\ image.
Unlike Fig. \ref{fig:coirac1},
both the IRAC and the CO images have been deprojected 
onto the galaxy plane and rotated (by 15$^\circ$) so that the major axis is
horizontal; units on $(x,y)$ axes ($\Delta$X$_G$/$\Delta$Y$_G$)
correspond to arcsec offsets along the major/minor axes with respect to the
AGN. 
The outer oval and stellar bar are marked by solid lines 
(in the deprojected rotated coordinate system).
The image field-of-view is $\sim$2.2\,kpc in diameter. %(r=1100~pc). 
The $\phi_i$-angles are measured from the $+$X
axis in the counter-clockwise direction. 
Right panel: {\bf b}) The same for \cotwo.  
\label{fig:dep-coirac1}}
\end{figure*}

The stellar potential in \nnn\ was derived from the IRAC 3.6\micron\ image by 
first rebinning to 0\farcs15 pixels, then deprojecting with
a PA\,=\,75$^\circ$ and an inclination of 30$^\circ$.
To account for the vertical mass distribution,
the rebinned deprojected image was convolved with an isothermal plane 
model of constant scale height $\sim$1/12th of the radial disk scalelength
\citep[e.g.,][]{quillen94,buta01}.
%Because we are interested only in {\it relative} changes in the mass distribution
%(see Sect. \ref{sec:torquederivation}), we invoke an arbitrary scaling constant to relate
%mass to light in the convolved image.
Figure \ref{fig:dep-coirac1} shows the deprojected IRAC image, 
with the \coone\ and \cotwo\ total intensity maps superimposed in
contours.
%This is proportional to the potential $\Phi$, but without the 
%convolution of the disk vertical mass distribution. 
The orientations of the stellar outer oval and the nuclear oval/bar are clearly visible
even in the deprojected image and are marked with a solid line.

%\begin{figure*}[!ht]

In the absence of extinction and stellar population gradients,
the mass-to-light ratio can be assumed to be constant across
this deprojected (face-on) galaxy image
\citep[e.g.,][]{freeman92,persic96};
the mass is therefore directly proportional to the 3.6\micron\ 
light distribution.
With this assumption, we derive the potential $\Phi$ by solving
the Poisson equation: $\Delta \Phi\,=\,4\,\pi\,G\,\rho$
using a Fast Fourier Transform (FFT) technique on a 256$\times$256
Cartesian grid as in \citet{combes90}.
Beyond a radius of 3.4\,kpc, $\rho$ is set to 0, thus suppressing any
spurious $m=4$ terms in the derived potential.
This is adequate to compute $\Phi$ in the central regions of the 
galaxy sampled by the PdB CO(1--0) primary beam.

\begin{figure*}
\centering
\includegraphics[width=0.35\linewidth,angle=-90, bb=56 35 297 720]{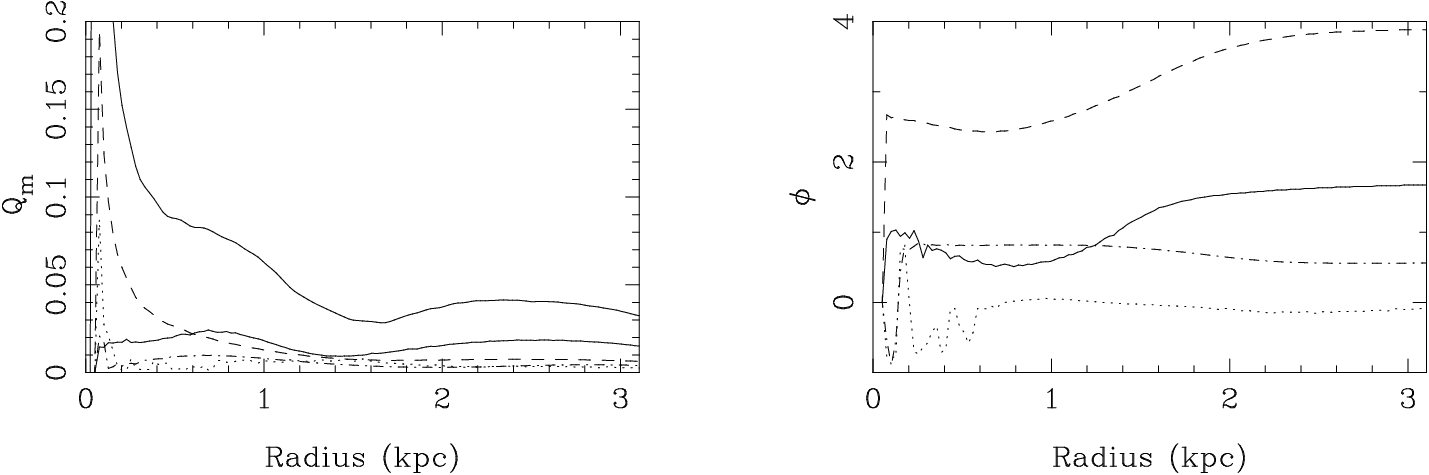}
\caption{Left panel: {\bf a}) The different normalised Fourier components $Q_m$ 
(defined in Eq. \ref{eqn:qm}, as the maximum tangential force, normalised to the 
radial force), for the potential of \nnn, derived from the NIR image. The top solid 
line is the total tangential force $Q_t$,
the next solid line is $Q_2$, the dashed line $Q_1$, the dotted line $Q_3$,
and the dotted-dashed one $Q_4$. 
Right panel: {\bf b}) The corresponding phases $\theta_m$ in radians, defined in Eq.
\ref{eqn:phi}, with the lines as in the left panel.
\label{fig:qplot}}
\end{figure*}

Following \citet{combes81}, we 
then expand the potential $\Phi$ in Fourier components:
\begin{equation}
\Phi(R,\theta)\,=\,\Phi_0(R) + \sum_m \Phi_m(R) \cos[m(\theta-\theta_m)]
\label{eqn:phi}
\end{equation}
and use a polar
Fourier transform method to isolate
the various angular modes $\Phi_m$.
It is convenient to represent each $m$-Fourier amplitude 
by the normalized ratio $Q_m(R)$: 
\begin{eqnarray}
Q_m(R)& = &\frac{m}{R} \Phi_m(R) / \left( {\partial \Phi_0(R)\over \partial R } \right) \\ \nonumber
 & = &m \Phi_m / R | F_0(R) | 
\label{eqn:qm}
\end{eqnarray}
which is the amplitude of the $m^{\rm th}$ harmonic of the
force relative to the total axisymmetric component.
These are essentially the harmonics of the angular derivatives of the potential,
expressed as normalized Fourier amplitudes. 
Figure \ref{fig:qplot} gives the radial run of the highest-order $Q_m(R)$ coefficients.
The nuclear (R$\simeq$1\,kpc) and the primary (R$\simeq$2.6\,kpc) bars
are clearly evident as local maxima (see left panel); 
moreover, the phase angle changes as would be
expected for the $\sim90^\circ$ transition from the nuclear to the
primary bar ($Q_2$ shown as a solid line, right panel).

\subsection{Derivation of the torques \label{sec:torquederivation}}

The angular derivative of the potential $\Phi(R,\theta)$ gives at each 
location the torque field 
per unit mass $\tau(x,y)=x F_y - y F_x$; by definition $\tau(x,y)$ is independent of
the gas distribution in the plane of the galaxy. 
The azimuthal average of $\tau$, using $N(x,y)$ as the weighting function, 
corresponds to the
global variation of the specific gas angular momentum transfer occurring at this
radius ($dL_s(x,y)/dt |_\theta$).
The $\theta$ subscript indicates an azimuthal average, evaluated at a given
radius $R(x,y)$.
For convenience of calculation on a pixelized image, we express the torques in 
Cartesian coordinates. 

\begin{figure*}
\centering
\includegraphics[width=0.9\linewidth]{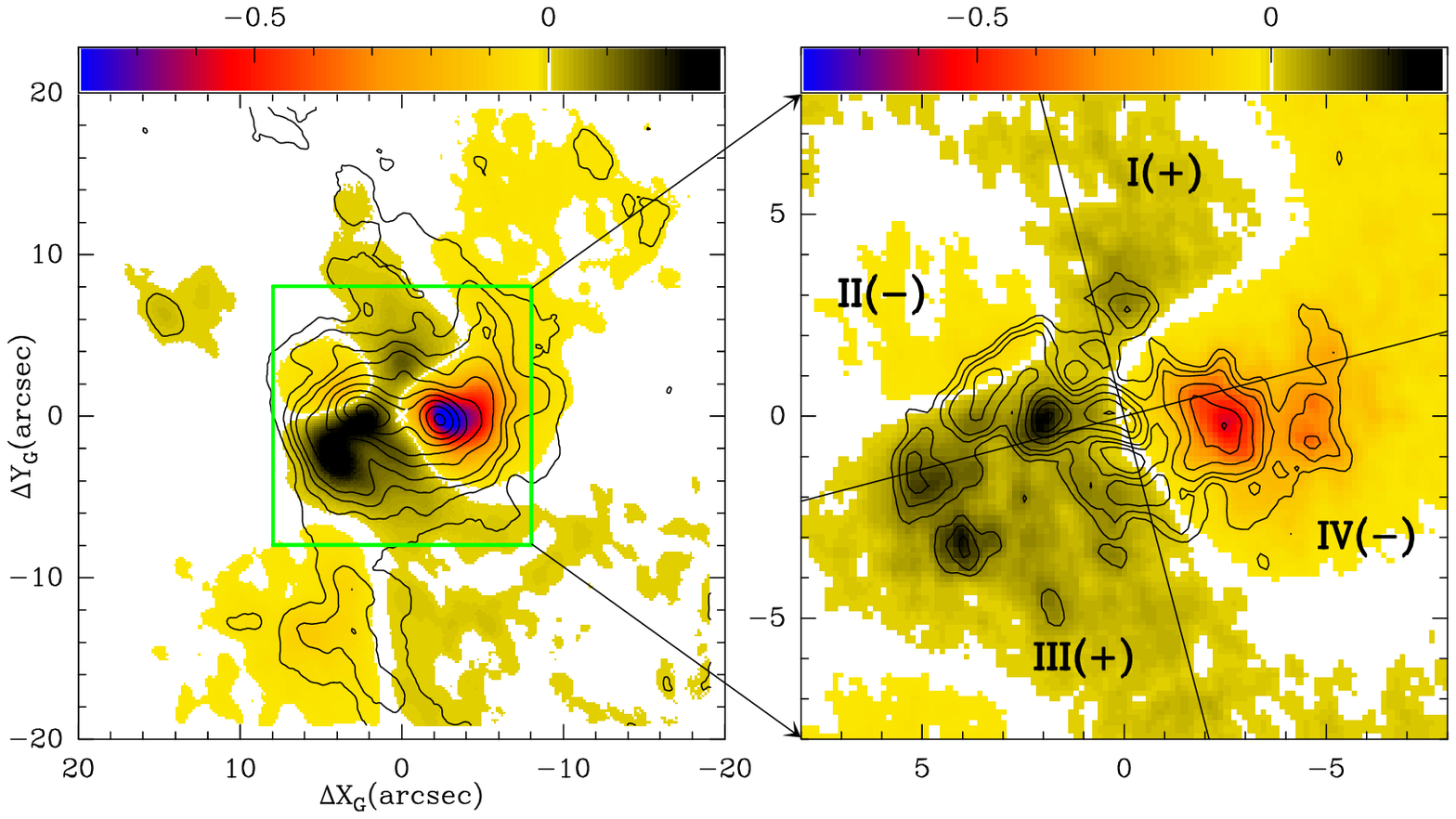}
\caption{Left panel: {\bf a}) \coone\ total intensity contoured over a false-color
representation of the torque maps $\tau$ derived as described in the text.
As in Fig. \ref{fig:dep-coirac1}, this is the deprojected representation, with
units on $(x,y)$ axes ($\Delta$X$_G$/$\Delta$Y$_G$)
corresponding to arcsec offsets along the major/minor axes with respect to the
AGN. 
The torque exerted on the spiral arm to the south is clearly associated with
the outer oval.
Right panel: {\bf b}) The same for \cotwo.  
Both panels show the typical butterfly diagram due to an $m=2$ perturbation,
and in {\bf b)} we have marked the
four quadrants with their positive (I, III) and negative torques (II, IV)
which drive the gas outward and inward, respectively.
As described in the text, the sign of the torques is determined by the rotation
direction of the gas, namely clockwise in \nnn.
\label{fig:dep-torques}}
\end{figure*}

The rotation direction of the gas in the galaxy disk determines the sign of
$\tau(x,y)$: positive (negative) if the torque 
drives the gas outward (inward) at $(x,y)$.
To calculate the effective variations of angular momentum in the galaxy plane,
we then weight the torques by the gas column densities $N(x,y)$ derived from the
\coone\ and 2--1 lines: $\tau(x,y) N(x,y)$. 
Fig. \ref{fig:dep-torques} shows the normalized version of these maps
$\tau^\prime$ where we have divided $\tau(x,y) N(x,y)$
by the azimuthal average of the gas density.
The $^\prime$ superscript indicates this normalized $\tau$, so that we can
use dimensionless quantities (both in Cartesian coordinates and below in polar ones).
Both panels in Fig. \ref{fig:dep-torques}
show the typical ``butterfly'' diagram due to an $m=2$ perturbation
\citep[e.g.][]{buta01,laurikainen02}.

%{\tt i'm not clear about these units; if they're
%normalized as in paper IV, we would have a maximum of unity. are they
%normalized but logarithmic? }
%}

%\begin{figure*}[!ht]

\begin{figure}
\centering
\includegraphics[width=0.9\linewidth]{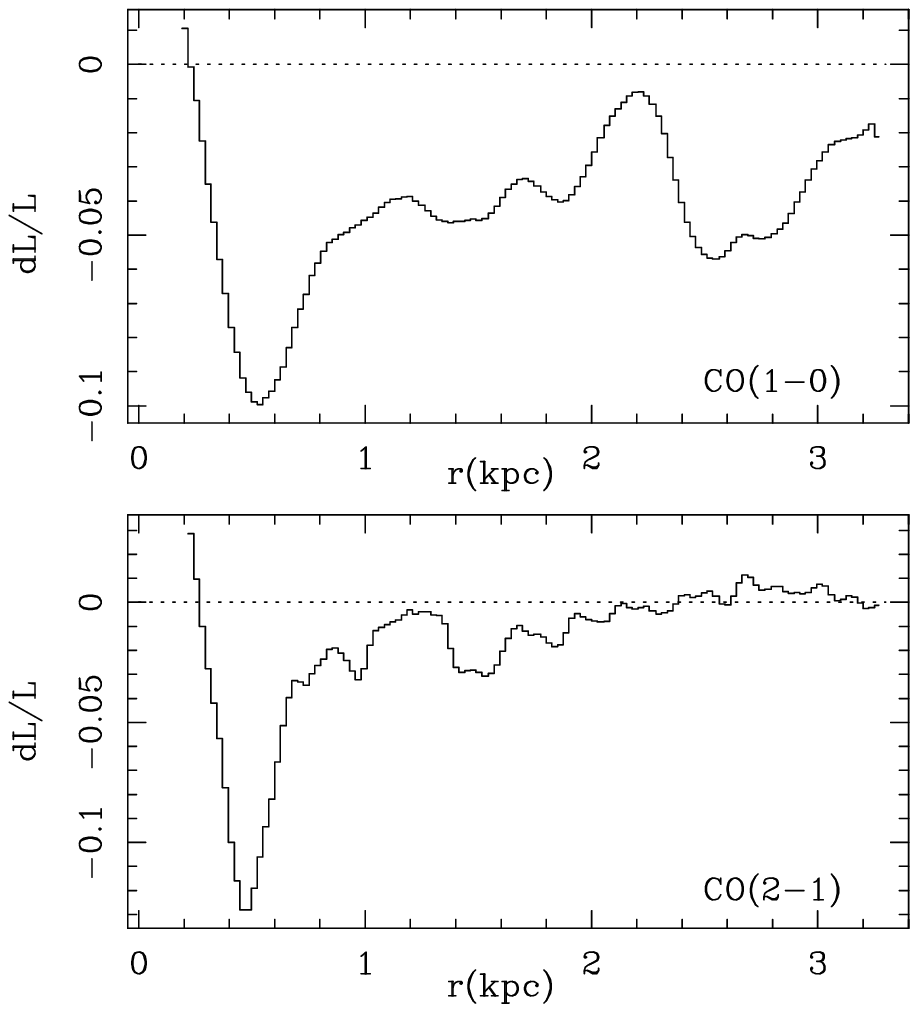}
\caption{Top panel: {\bf a}) The relative torque in dimensionless units
of $dL/L$ 
derived using the potential obtained from the \spitzer /IRAC 3.6\micron\ image
(1.2\arcsec\ pixels, undersampled beam).
Bottom panel: {\bf b}) The same for \cotwo.  
The \cotwo\ emission is less well sampled than \coone\ 
beyond a radius of $\sim$1.9\,kpc, 
because of the smaller beam size (FWHP$\sim$22\arcsec).
Despite the relatively poor
resolution of the gravitational potential image, 
both CO transitions give similar results; 
we see {\it systematically negative average torques within the entire
region probed by our observations}.
\label{fig:torques} 
\label{fig:radial-torques}}
\end{figure}

The radial gas flow induced by the torques is estimated by azimuthally averaging the
effective variations of angular momentum density in the plane:
\begin{equation}
\tau^\prime(R) = \frac{\int_\theta N(x,y)\times(x~F_y -y~F_x)}{\int_\theta N(x,y)}
\end{equation}
$\tau^\prime(R)$, by definition, represents the azimuthally-averaged time 
derivative of the 
specific angular momentum $L_s$. 
Azimuthal averages are derived with a radial binning commensurate with the 1\farcs2
resolution of the IRAC image.
Similarly to the torque maps (e.g., Fig. \ref{fig:dep-torques}), the sign of $\tau^\prime(R)$
defines whether the gas is subject to a net gain ($+$) or loss ($-$) of angular momentum.
Specifically, we assess the efficiency of the AGN fueling by estimating the average
fraction of the gas angular momentum transferred in one rotation period by the stellar potential.
This is done as a function of radius by defining the dimensionless function
$\Delta L/L$: 

\begin{equation}
{\Delta L\over L}=\left.{dL\over dt}~\right\vert_\theta\times \left.{1\over L}~\right\vert_\theta\times 
T_{rot}={\tau^\prime(R)\over L_\theta}\times T_{rot}
\label{tgrav}
\end{equation}
\noindent
where $T_{rot}$ is the rotation period, and
$L_\theta$ is assumed to be well represented by its azimuthal average, 
i.e., $L_\theta=R\times v_{rot}$. 
The absolute value of $L/\Delta L$ dictates how much time the stellar potential will
need to transfer the entirety of the total gas angular momentum.
Assuming that the gas response to the stellar potential is stationary
with respect to the potential reference frame during a few rotation periods, a small
value of $\Delta L / L$ implies that the stellar potential is inefficient at present. 
Figure \ref{fig:radial-torques} shows the $\Delta L/L$ relative torques derived
from CO(1--0) (top panel) and 2--1 (bottom).
Because of the limited resolution in the IRAC image (1 pixel$\simeq$200\,pc), we have 
masked the inner pixel. 
Beyond a radius of $\sim$1.9\,kpc, the \cotwo\ emission is 
less well sampled than \coone\ because of the smaller beam size ($\sim$22\arcsec).
The two CO lines are thus complementary, as \coone\ effectively traces the torques
beyond this radius, but at a more moderate spatial resolution than \cotwo.
It is evident that both lines give similar results, in the sense
that we see {\it systematically negative average torques in the inner few kpc.}

%\begin{figure}[!ht]

\section{Understanding the torques in \nnn \label{sec:understanding}}

The other galaxies in the NUGA sample analyzed so far
\citep{santi03,francoise04,santi05,fred07} mainly show positive 
torques near the nucleus, with negative torques at varying
amplitudes only for radii $\ga$500\,pc.
Only in NGC\,4579 are the average torques negative
down to 200\,pc, but the spatial resolution of the torque map 
for that galaxy is sufficiently good that 
we can see that, inside that radius,
the gas is flowing {\it away} from the nucleus.
\nnn\ is the first galaxy in our sample that shows systematically
negative torques down to the resolution limit imposed by
our torque image.
Moreover, in the entire region probed by our CO maps,
nowhere is the gas subjected to average positive torques.
The relatively poor resolution (1\farcs2) of the potential map
(IRAC 3.6\,\micron) makes it impossible to assess the torque amplitude
inside a radius of 200\,pc\footnote{As in \citet{santi05}, we attempted
to use the \hst\ WFPC2 F814W image corrected for extinction.
However, in \nnn, the extinction correction proved to be unreliable,
and we were forced to rely only on the \spitzer/IRAC image.}.
Indeed, the slightly positive torques in Fig. \ref{fig:radial-torques}
are unreliable, and may simply be an artefact of insufficient
spatial resolution.
Nevertheless, the systematically negative trend of the average torques
in \nnn\ hints that gas is being transported tantalizingly close to the
AGN.
This gas is certainly fuelling the nuclear starburst, and may, 
subsequently, be feeding the AGN.
In the following, we develop a dynamical scenario which will explain
why we think this occurs in \nnn.

\subsection{Main features \label{sec:features}}

Although the outer parts of NGC\,2782 show
clear signs of a past interaction (tidal tails, distorted disk, stellar sheets), 
the inner parts appear dynamically relaxed, not unexpectedly since
the dynamical timescale is shorter 
by one or two orders of magnitude at r$\sim$ 1kpc,
than at r$\sim$ 20kpc.

The previous sections have shown that \nnn\ contains
many features of a double-barred galaxy. 
Although the nuclear bar is unfortunately almost aligned with the galaxy
line of nodes, the disk is nearly 
face-on (30$^\circ$), which helps to distinguish the morphological
features.
In the near-infrared, there is
a nuclear stellar bar of radius $\sim$ 1\,kpc, 
embedded within an outer stellar oval of radius $\sim$ 2.5\,kpc.
This outer oval, which we will call the primary bar,
is oriented at PA $\sim$10$^\circ$, misaligned (and not quite perpendicular) 
with the nuclear stellar bar at PA$\sim$88$^\circ$.
The molecular gas distribution is mainly elongated along the nuclear bar,
and reveals two spiral features emerging from its ends, 
aligned with the conspicuous dust lanes already noted by \citet{jogee98,jogee99}.
These dust lanes are strikingly straight, reminiscent 
of barred galaxies with dust lanes parallel to the bar,
but located on the bar's leading edge on either side of the nucleus.
We therefore interpret the large-scale CO(1--0) features corresponding to the 
straight dust lanes as the gas response to the primary bar; 
they have indeed a similar orientation.
The inner elongated molecular morphology, evident in both CO(1--0)
and CO(2--1), constitutes the gas response to the nuclear bar.

The \ha\ morphology is more puzzling, because of
a clumpy dust obscuration which masks part of it. 
As noted by \citet{jogee98}, there are two main features in \ha:
one is the outflow driven by the starburst, perpendicular to
the nuclear stellar bar, and aligned with the radio continuum outflow
\citep{saikia94}, and the other is the elongated ring,
of about 1\,kpc in radius, encircling the nuclear bar. The ring
is particularly demarcated on the northern side, where it forms a 
180$^\circ$ arc, making the structure resemble more a
half-ring or what we called the ``cap of the mushroom'' in Sect.
\ref{sec:dust}. But there is also a southern equivalent (dubbed the
``mushroom stem''), largely obscured by dust. 
Toward the west, there is a remarkable dust finger,
devouring the ring, which superposes quite well with the CO distribution
(see Fig. \ref{fig:cohstvi}).

The two different \ha\ components are well distinguished by
their different kinematics: large outflowing velocities in the 
north-south component (bubble to the south in front of the dust lane, and 
obscured flow in the north behind it),
and a normally rotating ring around the nuclear bar \citep{yoshida99}.
We interpret the low-velocity rotating component as
the nuclear ring, associated with the inner Lindblad resonance (ILR) of
the primary bar, and encircling the nuclear bar, i.e. close
to the corotation (CR) of the nuclear bar (more exactly at the UHR, or 4:1 resonance).
The dust features are more conspicuous toward
the north because the near side is the north side.
The bulge is more evident toward the south,
masking the dust obscuration from the disk's far side.
But this does not imply that there is no dust and gas
there, obscuring the southern half-ring in \ha: on the contrary,
there is CO emission in this southern part, coinciding with
holes in the \ha\ distribution.

The \ha\ emission is the most intense in the partial ring, which lends support
to our ILR interpretation, since it is well-known
that resonances are locations where the gas is transiently stalled
due to the torques provided by the stellar bar. Thus, the gas has time to
form young stars, as typically revealed in many barred galaxies,
such as  NGC\,2997, NGC\,4314, or NGC\,4321 \citep{maoz96,benedict02,allard06}.
In the case of \nnn, because of the secondary bar,
gas is then driven further inward, as suggested by the CO
straight spiral arms (pitch angle close to 90$^\circ$) aligned
with the nuclear bar. This gas also forms stars, explaining the presence 
of \ha\ also in the nuclear disk.

\subsection{Dynamical interpretation \label{sec:interpretation}}

The dynamical scenario we propose to explain these main features is the following:
first a primary bar forms, most likely triggered by the interaction that gave rise
to the disturbed morphology in the outer parts of \nnn. 
The gravity torques from this primary bar drive the gas inward, and 
create a gaseous nuclear ring at its ILR. Intense star formation
gives rise to the H$\alpha$ nuclear ring, seen now as a sharp half-ring arc.
In the meantime, the gas inflow has weakened the primary bar 
\citep[e.g.,][]{bournaud02,bournaud05},
and a nuclear bar has dynamically decoupled, rotating at a different pattern speed.
The two different pattern speeds are usually such that the corotation
of the nuclear bar corresponds to the ILR of the primary bar,
thus avoiding excessive chaos in the orbits \citep{friedli93}.
The nuclear bar then governs the gas inflow toward the center.
Instead of being stalled at the primary ILR, now the gas is subject
to the negative gravity torques of the secondary bar,
between its CR and its own ILR. This explains why we see
CO emission inside the ILR of the primary bar, and why the torques
computation reveals that gas is flowing inward toward the center.

This scenario is slightly different from one where the gas decouples from 
the stars to form a gaseous nuclear bar with a faster \citep{englmaier04} or
slower \citep{heller01} pattern speed than the primary bar. We propose
here that the nuclear bar exists both in the stellar and gaseous components,
and always with a higher pattern speed than the primary.

\subsection{Numerical techniques \label{sec:code}}

To better understand the physical mechanisms, and 
reproduce the observed morphology, we
performed N-body simulations with stars and gas, including star formation.
We believe that the nuclear bar decoupling is a disk
phenomenon, occurring on short timescales, with little
contribution of a slowly rotating spherical component,
such as a bulge or dark matter halo, so we adopt the
simplification of 2D simulations, with the spherical components
added analytically.
Self-gravity is included only for the disk (gas$+$stars).

The N-body simulations were carried out
using the FFT algorithm to solve the Poisson equation, with
two kinds of grids: Cartesian and polar,
the latter optimizing the spatial resolution toward the center.
The useful Cartesian grid was 256$\times$256 (or 512$\times$512 in total
to avoid periodic images), with a total radius of 8\,kpc. The cell size
is then 62\,pc. The softening was taken as 250\,pc, in order to roughly
reproduce the actual effective softening from the thickness of the disk.
The polar grid is composed of NR$=80$ radial, and NT$=96$ azimuthal
separations. The radial spacing of the grid is exponential at large scales,
and linear at small scales, following the prescription by
\citet{pfenniger93}.
The polar-grid cell size ranges from 10\,pc at the center to 
$\sim$1\,kpc at the outskirts (at 22\,kpc).
The minimum softening is also fixed
at 250\,pc, and is larger in the outer regions.
The two different kinds of simulations, polar and Cartesian,
were performed separately, in order to gauge the effect of
different spatial resolution on the results;
both methods give similar outcomes.

The stellar component
is represented by 120000 particles, and the gas component by 40000.
Two rigid spherical potentials are added analytically, with a Plummer shape:
$$
\Phi_{b,h}(r) = - { {G M_{b,h}}\over {\sqrt{r^2 +r_{b,h}^2}} }
$$
where $M_{b}$ and  $r_{b}$ are the mass and characteristic radius of the
stellar spherical component (bulge),
and $M_{h}$ and  $r_{h}$ are the equivalent quantities for the
dark-matter halo.
These parameters are selected to fit the rotation curve of
\nnn, and are reported in Table \ref{tab:param-sim}.

The stellar disk is initially a Kuzmin-Toomre disk of surface density:
$$
\Sigma(r) = \Sigma_0 ( 1 +r^2/r_d^2 )^{-3/2}
$$
with mass M$_d$, $r_d$\,=\,3.25\,kpc, and truncated at 7\,kpc.
It is initially quite cold, with a Toomre Q parameter of 1.2.
The gas distribution is initially similar to the stellar one, with Q$_{gas}$ =1.
The time step is 0.1 Myr. The initial conditions of the runs
described here are given in Table \ref{tab:param-sim}.
The component subscripts refer to the bulge ("{\it b}"),
and the halo ("{\it h}") which are rigid,
and the disk ("{\it d}") which is ``live''.

\begin{table}[ht]
\caption[ ]{Initial conditions parameters}
\begin{flushleft}
\begin{tabular}{cccccc}  \hline
Run       & r$_b$ & M$_b$    & M$_d$    & M$_{h}$ & F$_{gas}$  \\
          & kpc  & M$_\odot$& M$_\odot$& M$_\odot$   &  \%        \\
\hline
Run A     &  0.2 &  2.3e10 &  1.4e11   & 4.5e10      &  10      \\
Run B     &  0.5 &  2.3e10 &  9.0e10   &  9.0e10     &  10       \\
Run C     &  0.2 &  2.3e10 &  1.4e11   & 4.5e10      &  5      \\
Run D     &  0.5 &  2.3e10 &  9.0e10   &  9.0e10     &  5       \\
\hline
\end{tabular}
\end{flushleft}
All masses are given inside 7\,kpc radius;
r$_d$ and  r$_{gas}$ are fixed at 3.25\,kpc, and r$_h$ at 7.5\,kpc.\\
The gas fraction F$_{gas}$ is the fraction of the total mass
(halo included).\\
\label{tab:param-sim}
\end{table}

The gas is treated as a self-gravitating component in the N-body
simulation, and its dissipation is treated by a sticky particle code,
as in \citet{combes85}. The initial gas-to-total mass ratio
(F$_{gas}$) ranges between 5 and 10\%. 
Star formation is included, with a conventional Schmidt-law recipe,
although this will not be discussed here, since the timescales
are short enough that the gas depletion is not significant. 
At the present time, the observed gas mass in \nnn\ within a $\sim$3\,kpc
radius is about $3\times10^9$ M$_\odot$\footnote{As derived from the
short-spacing single-dish observations of \citet{young95}, 
see Sect. \ref{sec:comorphology}.}.
The gas clouds are subject to inelastic collisions, with a collision
cell size of 67\,pc for the Cartesian code,
and 117\,pc for the polar code (region where particles are selected
to possibly collide). This corresponds to a lower limit for the average mean
free path of clouds between two collisions. The collisions are considered
every 2\,Myr. In a collision, the sign of the relative
cloud velocities is reversed and the absolute values are reduced:
relative velocities after the collision are only 0.65 times their original value.
The dissipation rate is controlled by this factor.
All gas particles have the same mass.

\begin{figure}[ht]
%\rotatebox{-90}{\includegraphics[width=0.75\linewidth]{../francoise2007/n2782_sim_vcur.ps}}
%{\includegraphics[width=\linewidth,bb=17 300 590 713]{n2782_omega.ps}}
{\includegraphics[width=\linewidth,bb=17 300 590 713]{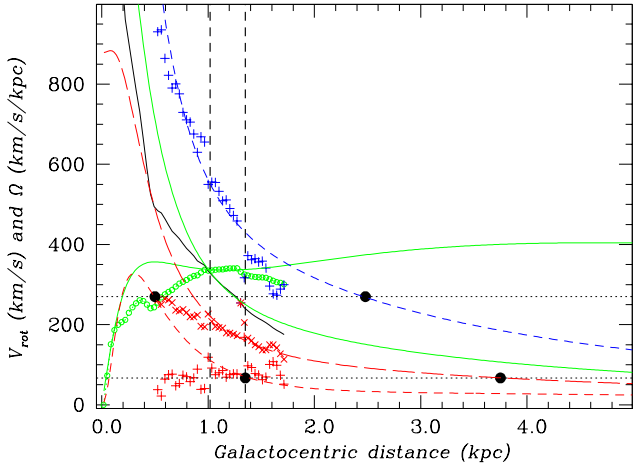}}
\caption{ Rotation and derived frequency curves $\Omega$,
$\Omega-\kappa/2$ and $\Omega+\kappa/2$ adopted
for the simulations (Run A), together with the empirical curves derived
from the CO(2--1) and (1--0) PV diagrams. 
The (green) open circles trace the empirical \vrot, coinciding with 
theoretical circular velocity \vcirc\ (traced by a solid line) 
at R\,=\,1\,kpc and at small radii.
The remaining solid lines are the empirical (black) and theoretical 
(green) $\Omega$ curves.
The long-dashed line is the $\Omega - \kappa/4$ curve,
and the short-dashed ones are the $\Omega \pm \kappa/2$ ones.
The corresponding data points are shown as $\times$ and $+$;
the data have been masked within 500\,pc to minimize confusion.
The two pattern speeds inferred from the simulations are shown as horizontal
dotted lines, and resonances are marked with filled circles.
The vertical dashed lines correspond to the nuclear bar length ($\sim$1\,kpc)
and the ILR of the outer oval ($\sim$1.3\,kpc), also roughly equal to the
nuclear bar's CR at $\sim$1.1\,kpc.
The apparent mismatch of the empirical and theoretical RCs,
in the range of 200$-$800\,pc, is due to non-circular 
streaming motions (Sect. \ref{sec:kinematics}) and asymmetry in the two sides
of the rotation curve (Sect. \ref{sec:rc})
which contribute to the inferred average \vrot.
} 
\label{fig:vcur}
\end{figure}

The rotation curve corresponding to Run A
is plotted in comparison to the empirical RC (see Sect. \ref{sec:rc}) 
in Fig.~\ref{fig:vcur}.
All other runs have similar rotation curves.
In the figure, we show the circular velocity \vcirc\ deduced from the potential
($d\Phi/dr\,=\,V_{\rm circ}^2/r$).
In the range of 200$-$800\,pc, there are non-circular 
streaming motions (see Sect. \ref{sec:kinematics})
which contribute to the derived empirical \vrot. 
Indeed, we have calculated \vrot\ of the gas from the model,
azimuthally averaged at a given radius, and find streaming motions of 
amplitude similar to the observed discrepancies between the empirical RC as 
traced by CO and the predicted \vcirc\footnote{As noted in Sect. \ref{sec:rc}, the
unfortunate coincidence of the PA of the bar and the disk line-of-nodes
makes streaming signatures difficult to diagnose in PV and isovelocity diagrams.}.
Given the uncertainties, and the asymmetry of the
two sides of the major axis (see Sect. \ref{sec:rc}), 
all theoretical curves are compatible with the empirical ones.
The data points for the epicyclic frequency resonances, marked by $\times$
and $+$ in Fig. \ref{fig:vcur}, are
calculated by numerically differentiating the smoothed empirical RC.
These data are also generally compatible
with the predictions of the numerical simulations.

\subsection{Simulation results \label{sec:simul}}

Several runs have been carried out in order to test the various parameters,
and evaluate the dependence of the results
on the central mass concentration and on the initial  gas
mass fraction. These parameters are crucial for the nuclear
bar decoupling.
Only four of these models are displayed in Table \ref{tab:param-sim}. 

We found that a high central mass concentration was
necessary to decouple the secondary bar, as already
shown by \citet{friedli93}.
This can be understood because
the two pattern speeds are close to the precessing rates
of the supporting orbits, which are traced by the $\Omega-\kappa/2$
curve. A high mass concentration produces a marked maximum
of this curve in the center, with a large radial gradient
favorable to the decoupling.
Also, a high gas mass fraction of at least 5\% of the total mass
is required for the decoupling. The nuclear bar is even
stronger with a gas fraction of 10\%.
In Run B and D, with a scale-length of the bulge twice that 
of Run A and C, the mass concentration was not enough to 
produce the decoupling of the nuclear bar. 

\begin{figure*}
\centering
%\rotatebox{0}{\includegraphics[width=0.7\linewidth]{../francoise2007/SG345-r4.ps}}
%\rotatebox{0}{\includegraphics[width=1.0\linewidth,angle=-90]{SGG345-r42.ps}}
%\rotatebox{0}{\includegraphics[width=1.0\linewidth,angle=-90]{8874fig17.ps}}
\rotatebox{0}{\includegraphics[width=1.0\linewidth]{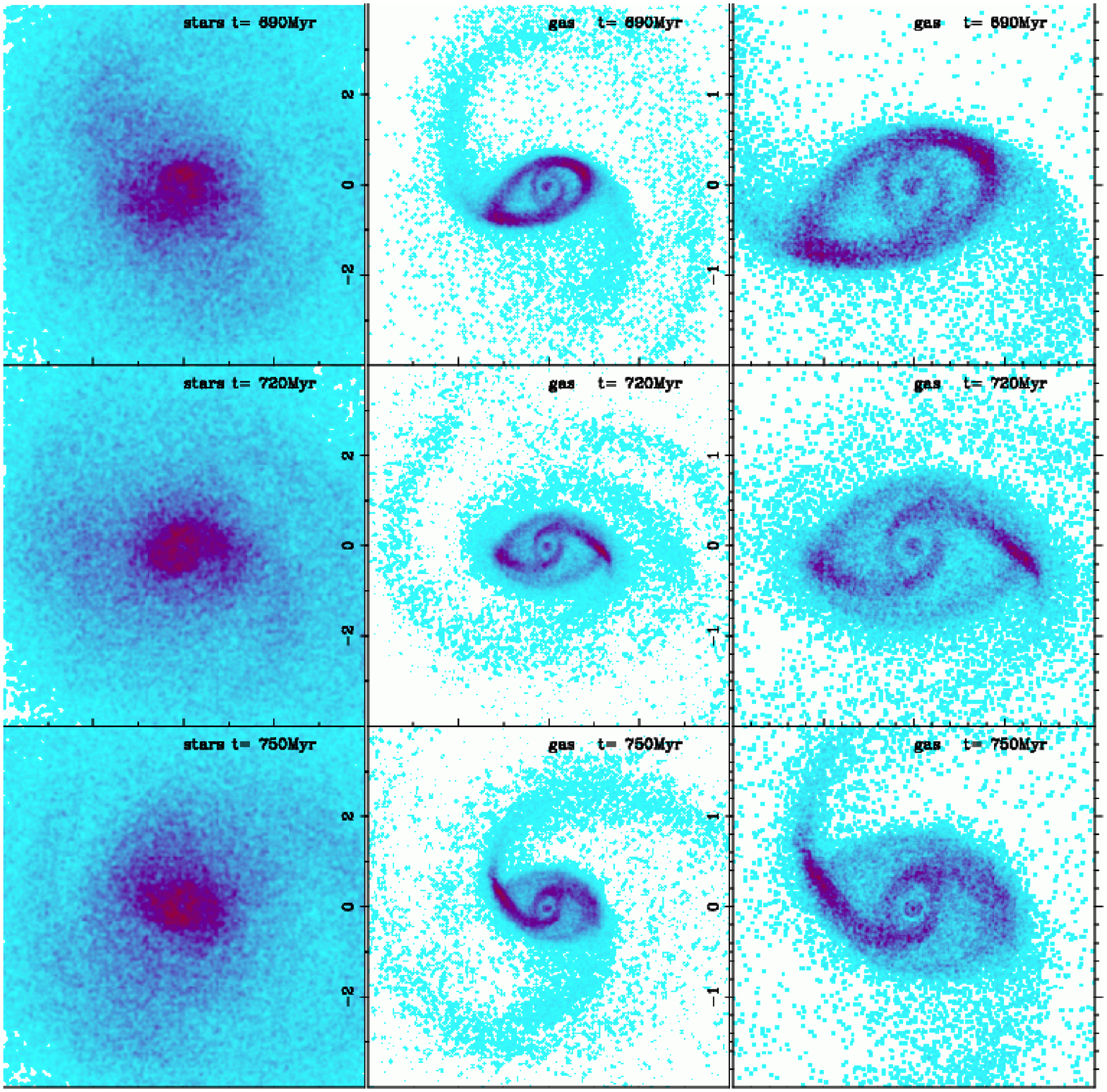}}
\caption{ {\bf Left panel:} False-color plot of the stellar component
distribution, in Run A (the intensity scale is linear and axes are in kpc).
{\bf Middle:}  False-color plot of the corresponding gaseous component
(the intensity scale is logarithmic).
{\bf Right:} Same as the middle panel, but with an expanded spatial scale.
It can be seen that the gas which was predominantly in the more external ring
(ILR of the primary bar) at T\,$=$\,690\,Myr, is falling progressively inward,
and is primarily inside the ILR at T\,$=$\,750\,Myr. The gas infalling
along the ``dust lanes'' is aligned along the nuclear bar, where the CO is 
observed in \nnn.
\label{fig:SG345}}
\end{figure*}

Run A, with a small bulge scalelength, and a high gas 
mass fraction, provides the best fit for \nnn.
The stellar and gaseous morphologies are displayed in 
Fig. \ref{fig:SG345} at three epochs, shortly after the second
bar decoupling. 
As seen in Fig. \ref{fig:SG345},
between $\sim$700 and 750\,Myr the gas is inflowing. Moreover,
the gas is always ``leading'' the stars, and thus is slowed down by
them.

\begin{figure*}
%\rotatebox{-90}{\includegraphics[width=0.37\linewidth]{newpower-run4p.ps}}
\rotatebox{-90}{\includegraphics[width=0.37\linewidth]{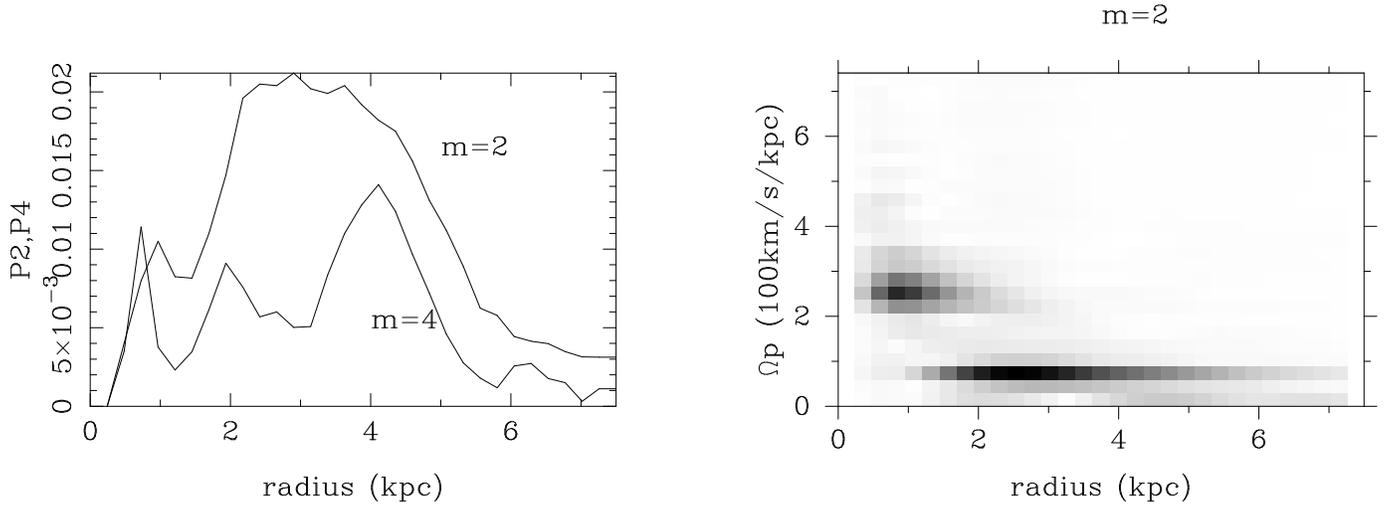}}
\caption{ {\bf Left:} Intensity of the $m=2$ and $m=4$ 
Fourier components of the potential in Run A, during
the second bar decoupling. 
{\bf Right:}  Power spectrum of the $m=2$ Fourier component,
giving the pattern speed as a function of radius, during a
period of 100\,Myr corresponding to the coexistence of the two bars.
Since the pattern speed of the nuclear bar varies with time
during this period, its $\Omega_b$ signature is somewhat widened.}
\label{fig:power-run4p}
\end{figure*}

The Fourier analysis of the gravitational potential
during the decoupling of the two bars clearly shows the development
of the two modes. 
The Fourier analysis  has been performed as a function of radius, 
every 2\,Myr, during 100\,Myr in this period. 
The Fourier transform of the results over
the time axis gives the pattern
speed as a function of radius, as shown in Fig \ref{fig:power-run4p},
with a resolution in $\Omega_p$ of 30\,\kmskpc.
Of necessity, the temporal averaging must
be performed over the lifetime of the transient features;
this averaging ``thickens'' the value of $\Omega_p$ in the plot
because of the time variation of the pattern speed of the nuclear bar.

Figure \ref{fig:power-run4p} lets us unambiguously isolate 
two bars in the potential, dominated by the stellar component:

\begin{enumerate}[(a)]
\item The nuclear bar with $\Omega_b\sim$270\,\kmskpc. 
Its corotation is at about 1.1\,kpc, at a radius just 
slightly larger than its size of 1\,kpc.
\item
The primary bar with $\Omega_b\sim$ 65\,\kmskpc\ and 
an ILR at $\sim$1.3\,kpc, roughly coincident with corotation of the nuclear bar;
the primary bar CR is at $\sim$5\,kpc, and its UHR at $\sim$3.6--3.8\,kpc. 
In the gas morphology, there is a (pseudo) inner ring  coinciding
with the UHR resonance (see Fig. \ref{fig:SG345}).
\end{enumerate}

From these identifications, we can interpret the
elongated nuclear ring as the ILR of the primary bar 
at R $\sim$ 1\,kpc, coinciding with the UHR/CR of the nuclear bar. 
This ring corresponds to the \ha\ arc, or half-ring, encircling
the NIR nuclear stellar bar.
%An analogous structure within the nuclear bar.
Inside the ILR ring (of the primary bar), the gas morphology reveals
two spiral arms with large pitch angle, now aligned with the nuclear bar.
These gas arms are leading with respect 
to the stellar bar, and this phase shift means that the
gravity torques from the stars are negative, driving
the gas inward. These arms are winding up in a ring
around the center, which corresponds to the ILR of the
nuclear bar. Indeed, it can be seen from the frequency
curves in Fig \ref{fig:vcur} that the pattern speed of the secondary
bar (270\,\kmskpc) crosses the $\Omega-\kappa/2$ curve at
about 300\,pc\footnote{Indeed, the high maximum of the $\Omega-\kappa/2$ 
curve is certainly
overestimated by the epicyclic approximation, since the orbits are
very elongated. 
Hence, although in our simulations, $\Omega_p$ of the
secondary bar would indicate two ILRs, it is more likely that there is only 
a single ILR, roughly in the middle at $\sim$200\,pc.}.
The two arms aligned along the nuclear bar can be identified
with the CO features, observed with the interferometer.
There is also a hole in the center in the gas distribution,
which could correspond to the ILR of the nuclear bar.

\subsection{Decoupled bars: gas infall in NGC\,2782 \label{sec:infall}}

The previous simulations, adapted to the rotation curve
of \nnn, support the scenario proposed in section \ref{sec:features}. 
A primary bar, perhaps triggered by the past interaction, has driven gas 
toward its ILR, and while this bar was weakening, a secondary embedded bar 
has decoupled,
prolonging the gas inflow inside the ILR ring, which is now
conspicuous in star formation and H$\alpha$ emission. The gas has
been driven into two spiral arms, aligned with the nuclear bar,
and is now winding up to the ILR of the nuclear bar, i.e. at
about 200--300\,pc radius. 
This scenario allows a coherent interpretation of the low-velocity H$\alpha$
component and the entire CO morphology, as well as of our finding of
negative gravity torques. 
In \nnn, there seems to be a ``smoking gun'', namely 
the presence of molecular gas inside the ILR 
of the primary bar, made possible by the decoupling of a secondary nuclear
bar from the primary one.
The gas there is certainly fueling
the central starburst, at the origin of the ionized gas bubble,
and in a second step should fuel directly the AGN.
The timescales in this central region are
quite short, about 6\,Myr for the gas rotation period at 300\,pc. 
Dynamical friction and viscous torques can then relatively
quickly provide gas to fuel the AGN \citep{santi05,santi07}, 
although the present simulations 
do not have enough spatial resolution to simulate that realistically.
Alternatively, the ILR of the nuclear bar suggested by our models
could impede further gas inflow inside the resolution limit of our images.

According to our simulations, there are two salient characteristics
of \nnn\ which lead to the secondary bar decoupling,
and the ensuing negative torques.
The first is its conspicuous stellar bulge, as suggested by the bulge-disk
decomposition described in Sect. \ref{sec:bd}.
This is the biggest contributor to the central mass concentration,
and the likely cause of the resonances that give rise to the gas
buildup in the rings, with the subsequent decoupling of the secondary
nuclear bar.
The second feature is the particularly high gas mass fraction in \nnn.
As discussed in Sect. \ref{sec:comorphology}, the mass of the
molecular gas in \nnn\ is the most extreme of all the NUGA galaxies
observed so far;
within a region $\sim$4\,kpc in diameter, the molecular gas
comprises 7\% of \mdyn.
This centrally concentrated molecular gas could be the result of
the interaction which \nnn\ has undergone. 
A high gas mass fraction is a necessary condition
for the decoupling of the secondary bar, and thus promotes infall.
Finally, we could be observing \nnn\ in a particularly favorable
epoch for the fueling of its AGN.
Future work on our statistical sample will help understand duty cycles
and timescales for the creation and maintenance of active accretion
in galactic nuclei.

\section{Summary and conclusions \label{sec:conclusions}}

We summarize our main results as follows:

\begin{enumerate}[(a)]
\item
High-resolution \twelveco\ observations of \nnn\ obtained with the PdB interferometer
show emission aligned with the stellar nuclear bar of radius $\sim$1\,kpc.
In \cotwo, we clearly resolve two spiral arms at high pitch angle along the nuclear bar;
in the \coone\ maps, the gas changes direction to form the beginning of two straight
dust lanes, aligned with an outer stellar oval reminiscent of a primary bar.
\item
From the stellar gravitational potential inferred from the IRAC 3.6\,\micron\ image
and the gas density from the \twelveco\ distributions,
we compute the azimuthally-averaged torques exerted by the stellar bars on the gas.
This calculation reveals
systematically negative torques in \nnn, down to the resolution limit of our image.
\item
The negative torques and the dynamics inferred from our observations are
explained by a scenario which is consistent with existing radio, optical,
and molecular-gas features in the circumnuclear region of the galaxy.
Star formation occurs in an \ha\ ring-like structure 
encircling the nuclear bar at the ILR of the primary bar/oval. 
The gas traced by CO emission is infalling to the center as a result
of gravity torques from the decoupled nuclear bar. 
The two high-pitch angle CO spiral arms are winding
up toward the center in a nuclear ring, corresponding to the 
ILR of the nuclear bar.
\item
This scenario is supported by N-body simulations which include
gas dissipation, and reproduce the secondary bar decoupling,
the formation of the elongated ring at the 1\,kpc-radius ILR of the primary bar,
and the gas inflow to the 200-300\,pc-radius ILR of the nuclear bar. 
\end{enumerate}

The average negative torques revealed by our analysis show that
infall to the central regions in \nnn\ is occurring in the present epoch.
Our numerical simulations suggest that the negative torques producing inflow are 
caused by the decoupling of the primary and nuclear bar.
This is made possible by the high central mass concentration and large gas
mass fraction in the galaxy, perhaps a result of the prior interaction
which caused the stellar sheets, ripples, and \hi\ tidal tails in its
outer regions. 
Future work will help understand what conditions are necessary for fueling
an AGN, and how the gas distribution, its kinematics, and the stellar gravitational 
potential interact to set the stage for gas inflow.

\begin{acknowledgements}
LKH sincerely thanks the LERMA-Observatoire de Paris for generous funding and
kind hospitality during the writing of this paper.
We would like to thank the anonymous referee for insightful questions which
helped clarify the text.
This research has made use of the NASA/IPAC Extragalactic Database (NED) which
is operated by the Jet Propulsion Laboratory, California Institute of
Technology, under contract with the National Aeronautics and Space
Administration.
\end{acknowledgements}

%\clearpage


\begin{thebibliography}{}
\bibitem[Allard et al.(2006)]{allard06} Allard, E.~L., Knapen, 
J.~H., Peletier, R.~F., \& Sarzi, M.\ 2006, \mnras, 371, 1087 
\bibitem[Benedict et al.(2002)]{benedict02} Benedict, G.~F., 
Howell, D.~A., J{\o}rgensen, I., Kenney, J.~D.~P., \& Smith, B.~J.\ 2002, 
\aj, 123, 1411 
\bibitem[Boone et al.(2007)]{fred07} Boone, F., Baker, A. J., Schinnerer, E.,
et al.\ 2007, \aap, 471, 113 (NUGA VII)
\bibitem[Boer et al.(1992)]{boer92} Boer, B., Schulz, H., \& Keel, W.~C.\ 1992, \aap, 260, 67 
\bibitem[Bournaud \& Combes(2002)]{bournaud02} Bournaud, F., \& 
Combes, F.\ 2002, \aap, 392, 83 
\bibitem[Bournaud et al.(2005)]{bournaud05} Bournaud, F., Combes, 
F., \& Semelin, B.\ 2005, \mnras, 364, L18 
\bibitem[Buta \& Block(2001)]{buta01} Buta, R., \& Block, 
D.~L.\ 2001, \apj, 550, 243 
\bibitem[Combes et al.(2004)]{francoise04} 
Combes, F., Garc{\'{\i}}a-Burillo, S., Boone, F., et al.\ 2004, \aap, 414, 857 (NUGA II)
\bibitem[Combes et al.(1990)]{combes90} Combes, F., Debbasch, 
F., Friedli, D., \& Pfenniger, D.\ 1990, \aap, 233, 82 
\bibitem[Combes \& Gerin(1985)]{combes85} Combes, F., \& Gerin, 
M.\ 1985, \aap, 150, 327 
\bibitem[Combes \& Sanders(1981)]{combes81} Combes, F., \& 
Sanders, R.~H.\ 1981, \aap, 96, 164 
\bibitem[de Jong(1996)]{dejong96} de Jong, R.~S.\ 1996, \aap, 313, 377 
%\bibitem[de Vaucouleurs et al.(1995)]{rc3} de Vaucouleurs, 
%G., de Vaucouleurs, A., Corwin, H.~G., Buta, R.~J., Paturel, G., \& Fouque, 
%P.\ 1995, VizieR Online Data Catalog, 7155, 0
%\bibitem[de Vaucouleurs et al.(1991)]{rc3} de Vaucouleurs, 
%G., de Vaucouleurs, A., Corwin, H.~G., Jr., Buta, R.~J., Paturel, G., \& 
%Fouque, P.\ 1991, Volume 1-3, XII, 2069 pp.~7 figs..~ Springer-Verlag 
%Berlin Heidelberg New York,   
\bibitem[Devereux(1989)]{devereux89} Devereux, N.~A.\ 1989, \apj, 346, 126 
\bibitem[Englmaier \& Shlosman(2004)]{englmaier04} Englmaier, P., 
\& Shlosman, I.\ 2004, \apjl, 617, L115 
\bibitem[Ferrarese et al.(2001)]{ferrarese01} Ferrarese, L., Pogge, 
R.~W., Peterson, B.~M., Merritt, D., Wandel, A., \& Joseph, C.~L.\ 2001, 
\apjl, 555, L79 
\bibitem[Freeman(1992)]{freeman92} Freeman, K.~C.\ 1992, Physics 
of Nearby Galaxies: Nature or Nurture?, 201 
\bibitem[Friedli \& Martinet(1993)]{friedli93} Friedli, D., \& 
Martinet, L.\ 1993, \aap, 277, 27 
\bibitem[Garc{\'{\i}}a-Burillo et al.(2003)]{santi03} 
Garc{\'{\i}}a-Burillo, S., Combes, F., Hunt, L. K., et al.\ 2003, \aap, 407, 485 (NUGA I)
\bibitem[Garc{\'{\i}}a-Burillo et al.(2005)]{santi05} 
Garc{\'{\i}}a-Burillo, S., Combes, F., Schinnerer, E., Boone, F., \& Hunt, 
L.~K.\ 2005, \aap, 441, 1011 (NUGA IV)
\bibitem[Garc{\'{\i}}a-Burillo et al.(2007)]{santi07} 
Garc{\'{\i}}a-Burillo, S., Combes, F., Usero, A., \& Graci{\'a}-Carpio, J.\ 
2007, New Astronomy Review, 51, 160
\bibitem[Heller et al.(2001)]{heller01} Heller, C., Shlosman, 
I., \& Englmaier, P.\ 2001, \apj, 553, 661 
\bibitem[Janiuk et al.(2004)]{janiuk04} Janiuk, A., Czerny, B., 
Siemiginowska, A., \& Szczerba, R.\ 2004, \apj, 602, 595 
\bibitem[Heckman et al.(2004)]{heckman04} Heckman, T.~M., 
Kauffmann, G., Brinchmann, J., Charlot, S., Tremonti, C., \& White, 
S.~D.~M.\ 2004, \apj, 613, 109 
\bibitem[Holtzman et al.(1995)]{holtzman95} Holtzman, J.~A., 
Burrows, C.~J., Casertano, S., Hester, J.~J., Trauger, J.~T., Watson, 
A.~M., \& Worthey, G.\ 1995, \pasp, 107, 1065 
\bibitem[Hopkins \& Hernquist(2006)]{hopkins06} Hopkins, P.~F., 
\& Hernquist, L.\ 2006, \apjs, 166, 1 
\bibitem[Hunt \& Giovanardi(1992)]{hunt92} Hunt, L.~K., \& 
Giovanardi, C.\ 1992, \aj, 104, 1018 
\bibitem[Hunt \& Malkan(1999)]{hunt99} Hunt, L.~K., \& Malkan, 
M.~A.\ 1999, \apj, 516, 660 
\bibitem[Hunt et al.(2004)]{hunt04} Hunt, L.~K., Pierini, D., 
\& Giovanardi, C.\ 2004, \aap, 414, 905 
\bibitem[Ishuzuki (1994)]{ishuzuki94}
Ishuzuki, S. 1994, in IAU Colloq. 140, 
Astronomy with Millimeter and Submillimeter Wave Interferometry, 
ed. M. Ishugiro \& J. M. Welch (ASP Conf. Ser. 59; San Francisco: ASP), 292
\bibitem[Jogee et al.(1998)]{jogee98} Jogee, S., Kenney, 
J.~D.~P., \& Smith, B.~J.\ 1998, \apjl, 494, L185 
\bibitem[Jogee et al.(1999)]{jogee99} Jogee, S., Kenney, 
J.~D.~P., \& Smith, B.~J.\ 1999, \apj, 526, 665 
\bibitem[Jogee et al.(2005)]{jogee05} Jogee, S., Scoville, N., 
\& Kenney, J.~D.~P.\ 2005, \apj, 630, 837 
\bibitem[Kent(1987)]{kent87} Kent, S.~M.\ 1987, \aj, 93, 816 
\bibitem[King \& Pringle(2007)]{king07} King, A.~R., \& 
Pringle, J.~E.\ 2007, MNRAS, in press ({\it astro-ph/0701679}) 
\bibitem[Knapen et al.(2002)]{knapen02} Knapen, J.~H., 
P{\'e}rez-Ram{\'{\i}}rez, D., \& Laine, S.\ 2002, \mnras, 337, 808 
\bibitem[Krips et al.(2005)]{melanie05} 
Krips, M., Eckart, A., Neri, R., et al.\ 2005, \aap, 442, 479 (NUGA III)
\bibitem[Krips et al.(2007)]{melanie07} Krips, M.,Eckart, A., 
Krichbaum, T. P., et al.\ 2007, \aap, 464, 553 (NUGA V) %in press ({\it astro-ph/0612772}) 
%\bibitem[Laurikainen et al.(2004)]{laurikainen04} Laurikainen, E., 
%Salo, H., \& Buta, R.\ 2004, \apj, 607, 103 
\bibitem[Laurikainen \& Salo(2002)]{laurikainen02} Laurikainen, E., 
\& Salo, H.\ 2002, \mnras, 337, 1118 
\bibitem[Lequeux(1983)]{lequeux83} Lequeux, J.\ 1983, \aap, 125, 394 
\bibitem[Makovoz \& Marleau(2005)]{makovoz05} Makovoz, D., \& 
Marleau, F.~R.\ 2005, \pasp, 117, 1113 
\bibitem[Maoz et al.(1996)]{maoz96} Maoz, D., Barth, A.~J., 
Sternberg, A., et al., \aj, 111, 2248 
%\bibitem[Marconi \& Hunt(2003)]{marconi03} Marconi, A., \& Hunt, 
%L.~K.\ 2003, \apjl, 589, L21 
\bibitem[Marecki et al.(2003)]{marecki03} Marecki, A., Spencer, 
R.~E., \& Kunert, M.\ 2003, Publications of the Astronomical Society of 
Australia, 20, 46 
%\bibitem[McElroy(1995)]{mcelroy95} McElroy, D.~B.\ 1995, \apjs, 100, 105 
\bibitem[Moriondo et al.(1998)]{moriondo98} Moriondo, G., 
Giovanardi, C., \& Hunt, L.~K.\ 1998, \aaps, 130, 81 
\bibitem[Mould et al.(2000)]{mould00} Mould, J.~R.,
Huchra, John P., Freedman, Wendy L., et al.\ 2000, \apj, 529, 786 
\bibitem[Mulchaey \& Regan(1997)]{mulchaey97} Mulchaey, J.~S., \& 
Regan, M.~W.\ 1997, \apjl, 482, L135 
\bibitem[Narayanan et al.(2006)]{narayanan06} Narayanan, D.,
Cox, T. J., Robertson, B., et al.\ 2006, \apjl, 642, L107 
\bibitem[Peng et al.(2002)]{peng02} Peng, C.~Y., Ho, L.~C., 
Impey, C.~D., \& Rix, H.-W.\ 2002, \aj, 124, 266 
\bibitem[Persic et al.(1996)]{persic96} Persic, M., Salucci, P., 
\& Stel, F.\ 1996, \mnras, 281, 27 
\bibitem[Pfenniger \& Friedli(1993)]{pfenniger93} Pfenniger, D., \& 
Friedli, D.\ 1993, \aap, 270, 561 
\bibitem[Quillen et al.(1994)]{quillen94} Quillen, A.~C., Frogel, 
J.~A., \& Gonzalez, R.~A.\ 1994, \apj, 437, 162 
\bibitem[Saikia et al.(1994)]{saikia94} Saikia, D.~J., Pedlar, 
A., Unger, S.~W., \& Axon, D.~J.\ 1994, \mnras, 270, 46 
\bibitem[Sakamoto et al.(1999)]{sakamoto99} Sakamoto, K., Okumura, 
S.~K., Ishizuki, S., \& Scoville, N.~Z.\ 1999, \apj, 525, 691 
\bibitem[Schweizer \& Seitzer(1988)]{schweizer88} Schweizer, F., \& 
Seitzer, P.\ 1988, \apj, 328, 88 
\bibitem[Smith(1991)]{smith91} Smith, B.~J.\ 1991, \apj, 378, 39 
\bibitem[Smith(1994)]{smith94} Smith, B.~J.\ 1994, \aj, 107, 1695 
\bibitem[Smith et al.(1999)]{smith99} Smith, B.~J., Struck, C., 
Kenney, J.~D.~P., \& Jogee, S.\ 1999, \aj, 117, 1237 
\bibitem[Solomon \& Barrett(1991)]{solomon91} Solomon, P.~M., \& 
Barrett, J.~W.\ 1991, Dynamics of Galaxies and Their Molecular Cloud 
Distributions, 146, 235 
%\bibitem[Terlevich et al.(1990)]{terlevich90} Terlevich, E., Diaz, 
%A.~I., \& Terlevich, R.\ 1990, \mnras, 242, 271 
%\bibitem[Tremaine et al.(2002)]{tremaine02} Tremaine, S., et al.\ 
%2002, \apj, 574, 740 
\bibitem[Yoshida et al.(1999)]{yoshida99} Yoshida, M., Taniguchi, 
Y., \& Murayama, T.\ 1999, \aj, 117, 1158 
\bibitem[Young et al.(1995)]{young95} Young, J.~S., Xie, S., 
Tacconi, L., et al.\ 1995, \apjs, 98, 219 
\bibitem[Zhang et al.(2006)]{zhang06} Zhang, J.~S., Henkel, C., 
Kadler, M., et al. \ 2006, \aap, 450, 933 
\end{thebibliography}
\end{document}